\documentclass[12pt]{article}
\usepackage{latexsym,amssymb,graphicx}
\newtheorem{Def1}{Definition}
\newtheorem{Lemma}[Def1]{Lemma}
\newtheorem{Prop}[Def1]{Proposition}
\newtheorem{Theorem}[Def1]{Theorem}
\newtheorem{Cor}[Def1]{Corollary}

\newtheorem{Claim}[Def1]{Claim}
\newtheorem{Problem}[Def1]{Problem}
\newtheorem{Ex1}[Def1]{Example}

\newenvironment{Def} {\begin{Def1} \begin{upshape}} {\end{upshape} \end{Def1}}
\newenvironment{Example} {\begin{Ex1} \begin{upshape}} {\end{upshape} \end{Ex1}}
\newenvironment{MyPar}[1] {\medskip\noindent \textbf{#1.} } {}
\newenvironment{Proof}[1][.]{\medskip\noindent \textbf{Proof}#1 }
  {\hspace*{0mm}\hfill $\Box$ \medskip}
\newenvironment{keywords}{\centerline{\bf\small
Keywords\vspace{-1.5ex}}\begin{quote}\small}{\par\end{quote}\vskip 1ex}

\def\baq#1\eaq{\begin{align}#1\end{align}}
\def\baqn#1\eaqn{\begin{align*}#1\end{align*}}
\def\beq#1\eeq{\begin{equation}#1\end{equation}}
\def\beqn#1\eeqn{\begin{displaymath}#1\end{displaymath}}
\def\bqa#1\eqa{\begin{eqnarray}#1\end{eqnarray}}
\def\bqan#1\eqan{\begin{eqnarray*}#1\end{eqnarray*}}

\def\calC{\mathcal C}

\def\calM{\mathcal M}
\def\calN{\mathcal N}

\def\calU{\mathcal U}

\def\calX{\mathcal X}

\def\NNN{\mathbb N}
\def\RRR{\mathbb R}
\def\QQQ{\mathbb Q}

\def\BBB{\{0,1\}}

\def\Expect{{\mathbf E}}
\def\Prob{{\mathbf P}}
\def\rrho{\varrho}
\def\eps{\varepsilon}
\def\epstr{\epsilon}

\def\equa{\stackrel+=}
\def\leqa{\stackrel+\leq}
\def\geqa{\stackrel+\geq}

\def\eqm{\stackrel\times=}
\def\leqm{\stackrel\times\leq}
\def\geqm{\stackrel\times\geq}

\def\leqt{_{1:t}}
\def\ltt{_{<t}}
\def\leqn{_{1:n}}

\def\leqss{_{1:s}}
\def\ltinf{_{<\infty}}

\def\_norm{_\mathrm{norm}}

\newcommand{\zwidths}[1]{\rlap{$\scriptstyle #1$}}

\def\for_all{\;\mbox{ for all }\;}
\def\such_that{\;\mbox{ such that }\;}
\def\wenn{\;\mbox{ if }\;}
\def\und{\;\mbox{ and }\;}

\def\lb{{\log_2}}                      % binary logarithm
\def\l{\ell}

\def\ph{\varphi}
\def\th{\vartheta}

\def\toinfty#1{\stackrel{#1\to\infty}{\longrightarrow}}
\def\toims{\stackrel{i.m.s.}{\longrightarrow}}

\def\tfrac#1#2{{\textstyle\frac{#1}{#2}}}

\def\K{K\!}

\def\M{M}

\def\Km{K\!m}

\def\tfrac#1#2{{\textstyle\frac{#1}{#2}}}

\begin{document}

\title{\normalsize\sc Technical Report \hfill IDSIA-13-05
\vskip 2mm\bf\LARGE\hrule height5pt \vskip 6mm
Asymptotics of Discrete MDL \\ for Online Prediction\thanks{
A shorter version of this paper \cite{Poland:04mdl} appeared in
COLT 2004.}
\vskip 6mm \hrule height2pt \vskip 5mm}
\date{6 June 2005}

\author{{\bf Jan Poland} and {\bf Marcus Hutter}\\
IDSIA, Galleria 2, CH-6928 Manno-Lugano, Switzerland\\
\texttt{\{jan,marcus\}@idsia.ch}\hspace{11ex}http://www.idsia.ch}

\maketitle

\vspace{-5ex}
\begin{abstract}\noindent
Minimum Description Length (MDL) is an important principle for
induction and prediction, with strong relations to optimal
Bayesian learning. This paper deals with learning non-i.i.d.\
processes by means of two-part MDL, where the underlying model
class is countable. We consider the online learning framework,
i.e.\ observations come in one by one, and the predictor is
allowed to update his state of mind after each time step. We
identify two ways of predicting by MDL for this setup, namely
a \emph{static} and a \emph{dynamic} one. (A third variant,
hybrid MDL, will turn out inferior.) We will prove that under
the only assumption that the data is generated by a
distribution contained in the model class, the MDL predictions
converge to the true values almost surely. This is
accomplished by proving finite bounds on the quadratic, the
Hellinger, and the Kullback-Leibler loss of the MDL learner,
which are however exponentially worse than for Bayesian
prediction. We demonstrate that
these bounds are sharp, even for model classes containing only
Bernoulli distributions. We show how these bounds imply regret
bounds for arbitrary loss functions. Our results apply to a
wide range of setups, namely sequence prediction, pattern
classification, regression, and universal induction in the
sense of Algorithmic Information Theory among others.
\end{abstract}


%%%%%%%%%%%%%%%%%%%%%%%%%%%%%%%%%%%%%%%%%%%%%%%%%%%%%%%%%%%%%%%
\section{Introduction}
%%%%%%%%%%%%%%%%%%%%%%%%%%%%%%%%%%%%%%%%%%%%%%%%%%%%%%%%%%%%%%%

``Always prefer the \emph{simplest} explanation for your
observation," says Occam's razor. In Learning and Information
Theory, simplicity is often quantified in terms of description
length, giving immediate rise to the Minimum Description
Length (MDL) principle
\cite{Wallace:68,Rissanen:78,Gruenwald:98}. Thus MDL can be
seen as a strategy against overfitting. An alternative way to
think of MDL is Bayesian. The explanations for the
observations (the \emph{models}) are endowed with a prior.
Then the model having maximum a posteriori (MAP) probability
is also a two-part MDL estimate, where the correspondence
between probabilities and description lengths is simply by a
negative logarithm.

How does two-part MDL perform for prediction? Some very accurate
answers to this question have been already given. If the data is
generated by an independently identically distributed (i.i.d.)
process, then the MDL estimates are consistent \cite{Barron:91}.
In this case, an important quantity to consider is the \emph{index
of resolvability}, which depends on the complexity of the data
generating process. This quantity is a tight bound on the regret
in terms of coding (i.e.\ the excess code length). Moreover, the
index of resolvability also bounds the predictive regret, namely
the rate of convergence of the predictive distribution to the true
one. These results apply to both discrete and continuously
parameterized model classes, where in the latter case the MDL
estimator must be discretized with an appropriate precision.

Under the relaxed assumption that the data generating process
obeys a central limit theorem and some additional conditions,
Rissanen \cite{Rissanen:96,Barron:98} proves an asymptotic
bound on the regret of MDL codes. Here, he also removes the
coding redundancy arising if two-part codes are defined in the
straightforward way. The resulting bound is very similar to
that in \cite{Clarke:90} for Bayes mixture codes and i.i.d.\
processes, where the i.i.d.\ assumption may also be relaxed
\cite{Hutter:03optisp}. Other similar and related results can
be found in \cite{Ghosal:01,Ghosal:04}.

In this work, we develop new methods in order to arrive at very
general consistency theorems for MDL on \emph{countable model
classes}. Our setup is \emph{online sequence prediction}, that
is, the symbols $x_1,x_2,\ldots$ of an infinite sequence
are revealed successively by the environment, where our task
is to predict the next symbol in each time step.
Consistency is established by proving \emph{finite cumulative
bounds} on the differences of the predictive to the true
distribution. Differences will be measured in terms of the
relative entropy, the quadratic distance, and the Hellinger
distance. Most of our results are based on the only assumption
that the data generating process is \emph{contained in the
models class}. (The discussion of how strong this assumption
is, will be postponed to the last section.) Our results imply
regret bounds with \emph{arbitrary} loss functions. Moreover,
they can be directly applied to important general setups such
as pattern classification, regression, and universal
induction.

As many scientific models (e.g.\ in physics or biology) are
smooth, much statistical work is focussed on continuous model
classes. On the other hand, the largest relevant classes from a
computational point of view are at most countable. In particular,
the field of Algorithmic Information Theory (also known as
Kolmogorov Complexity, e.g.\ \cite{Zvonkin:70,Li:97,Calude:02,Hutter:04uaibook})
studies the class of \emph{all lower-semicomputable semimeasures}.
Then there is a one-to-one correspondence of models and programs
on a fixed universal Turing Machine. (Since programs need not halt
on each input, models are semimeasures instead of measures, see
e.g.\ \cite{Li:97} for details). This model class can be
considered the largest one which can be in the limit processed
under standard computational restrictions. We will develop all our
results for semimeasures, so that they can be applied in this
context, which we refer to as \emph{universal} sequence
prediction.

In the universal setup, the Bayes mixture is also termed
Solomonoff-Levin prior and has been intensely studied first by
Solomonoff \cite{Solomonoff:64,Solomonoff:78}. Its predictive
properties are excellent \cite{Hutter:99errbnd,Hutter:04uaibook}.
Precisely one can bound the cumulative
loss by the complexity of the data generating process. This is the
reference performance we compare MDL to. It turns out that the
predictive properties of MDL can be exponentially worse, even in
the case that the model class contains only Bernoulli
distributions. Another related quantity in the universal setup is
\emph{one-part MDL}, which has been studied in
\cite{Hutter:03unimdl}. We will briefly encounter it in Section
\ref{subsecUniversal}.

The paper is structured as follows. Section \ref{secPN}
establishes basic definitions. In Section \ref{secMDL}, we
introduce the MDL estimator and show how it can be used for
sequence prediction in at least three ways. Sections
\ref{secDynamic} and \ref{secStatic} are devoted to convergence
theorems. In Sections \ref{secHybrid} and \ref{secStabilization},
we study the stabilization properties of the MDL estimator.
Section \ref{secApp} presents more general loss bounds as well as
three important applications: pattern classification, regression,
and universal induction. Finally, Section \ref{secDC} contains the
conclusions.

%%%%%%%%%%%%%%%%%%%%%%%%%%%%%%%%%%%%%%%%%%%%%%%%%%%%%%%%%%%%%%%
\section{Prerequisites and Notation}\label{secPN}
%%%%%%%%%%%%%%%%%%%%%%%%%%%%%%%%%%%%%%%%%%%%%%%%%%%%%%%%%%%%%%%

We build on the notation of \cite{Li:97} and
\cite{Hutter:04uaibook}. Let the alphabet $\calX$ be a finite set
of symbols. We consider the spaces $\calX^*$ and $\calX^\infty$ of
finite strings and infinite sequences over $\calX$. The initial
part of a sequence up to a time $t\in\NNN$ or $t-1\in\NNN$ is
denoted by
$x\leqt$ or $x\ltt$, respectively. The empty string is denoted by
$\epstr$.

A {\it semimeasure} is a function $\nu:\calX^*\to[0,1]$ such that
\beq
\label{eq:semimeasure}
\nu(\epstr)\leq 1 \und
\nu(x)\geq \sum_{a\in\calX}\nu(xa) \for_all x\in\calX^*
\eeq
holds. If equality holds in both inequalities of
(\ref{eq:semimeasure}), then we have a {\it measure}.
Intuitively, the quantity $\nu(x)$ can be understood as the
probability that a data generating process yields a string
starting with $x$. Then, for a measure, the probabilities of
all joint continuations of $x$ add up to $\nu(x)$, while for a
semimeasure, there may be a ``probability leak"
(\ref{eq:semimeasure}). Recall that we are interested in
semimeasures (and not only in measures) because of their
correspondence to programs on a fixed universal Turing machine
in the universal setup and our inability to decide the halting
problem.

Let $\calC$ be a countable class of (semi)measures, i.e.\
$\calC=\{\nu_i:i\in I\}$ with finite or infinite index set
$I\subseteq\NNN$. A (semi)measure $\tilde\nu$ {\it dominates} the
class $\calC$ iff for every $\nu_i\in\calC$ there is a constant
$c_i>0$ such that
$\nu(x)\geq c_i\cdot\nu_i(x)$ holds for all $x\in\calX^*$. A
dominant semimeasure $\tilde\nu$ need not be contained in $\calC$.

Each (semi)measure $\nu\in\calC$ is associated with a weight
$w_\nu>0$, and we require $\sum_{\nu} w_\nu\leq 1$. We may interpret the
weights as a {\it prior} on $\calC$. Then it is obvious that
the Bayes mixture
\beq
\label{eq:xi}
\xi(x)\equiv\xi_{[\calC]}(x):=\sum_{\nu\in\calC}w_\nu \nu(x)
\quad (\mbox{for }x\in\calX^*)
\eeq
dominates $\calC$. Assume that there is some measure
$\mu\in\calC$, the {\it true distribution}, generating sequences
$x\ltinf\in\calX^\infty$. Typically $\mu$ is unknown.
(Note that we require $\mu$ to be a measure: The data stream
always continues, there are no ``probability leaks".) If some
initial part $x\ltt$ of a sequence is given, the probability
of observing $x_t\in\calX$ as a next symbol is
\beq
  \label{eq:basicprediction}
  \mu(x_t|x\ltt)=\frac{\mu(x\ltt x_t)}{\mu(x\ltt)} \wenn
  \mu(x\ltt)>0 \ \und\  \mu(x_t|x\ltt)=0 \wenn \mu(x\ltt)=0.
\eeq
and, for well-definedness, $\mu(x_t|x\ltt)=0$ if $\mu(x\ltt)=0$ (this case
has probability zero). Note that $\mu(x_t|x\ltt)$ can depend on the
complete history $x\ltt$. We may generally define the quantity
(\ref{eq:basicprediction}) for {\it any} function
$\ph:\calX^*\to[0,1]$; we call $\ph(x_t|x\ltt):=\frac{\ph(x\leqt)}{\ph(x\ltt)}$ the {\it
$\ph$-prediction}. Clearly, this is not necessarily a probability
on $\calX$ for general $\ph$. For a semimeasure
$\nu$ in particular, the $\nu$-prediction $\nu(\cdot|x\ltt)$ is a
semimeasure on $\calX$.

We define the {\it expectation} with respect to the true
probability $\mu$: Let $n\geq 0$ and $f:\calX^n\to\RRR$ be a
function, then
\beq
\label{eq:expectation}
\Expect\ f=\Expect\ f(x\leqn)=\sum_{x\leqn\in\calX^n}\mu(x\leqn)f(x\leqn).
\eeq
More general, the expectation may be defined as an integral over
infinite sequences. But since we won't need it, we can keep
things simple. The following is a central result about
prediction with Bayes mixtures in a form independent of
Algorithmic Information Theory.

\begin{Theorem}
\label{Theorem:Solomonoff}
For any class of (semi)measures $\calC$ containing the true
distribution $\mu$, which is a measure, we have
\beq
\label{eq:Solomonoff}
\sum_{t=1}^\infty \Expect \sum_{a\in\calX}
  \Big( \mu(a|x\ltt)-\xi(a|x\ltt) \Big)^2 \ \leq \ \ln w_\mu^{-1}.
\eeq
\end{Theorem}

This was found by Solomonoff (\cite{Solomonoff:78}) for
universal sequence prediction. A proof is also given in
\cite{Li:97} (only for binary alphabet) or
\cite{Hutter:04uaibook} (arbitrary alphabet). It is
surprisingly simple once Lemma \ref{Lemma:EntropyIneq} is
known. A few lines analogous to
(\ref{Eq:TheoremRhoConvergesIMS0}) and
(\ref{Eq:TheoremRhoConvergesIMS00}) exploiting the dominance
of
$\xi$ are sufficient.

One should be aware that the condition $\mu\in\calC$ is
essential in general, for both Bayes and MDL predictions
\cite{Gruenwald:04}. On the other hand, one can show that if
there is an element in
$\calC$ which is sufficiently close to $\mu$ in an appropriate
sense, then still good predictive properties hold
\cite{Hutter:03optisp}.

Note that although $w_\nu$ can be interpreted as a prior on the
model class, we do not assume any probabilistic structure for
$\calC$ (e.g.\ a sampling mechanism). The theorem rather states
that the cumulative loss is bounded by a quantity depending on the
complexity $\ln w_\mu^{-1}$ of the true distribution. The same
kind of assertion will be proven for MDL predictions later.

The bound (\ref{eq:Solomonoff}) implies that the
$\xi$-predictions converge to the $\mu$-predictions almost surely
(i.e.\ with $\mu$-probability one). This is not hard to see, since
with the abbreviation
$s_t=\sum_a\big( \mu(a|x\ltt)-\xi(a|x\ltt) \big)^2$ and for each
$\eps>0$, we have
\bqa
\nonumber
\Prob\Big(\exists t\geq n: s_t\geq\eps\Big) & = &
\Prob\Big(\bigcup_{t\geq n}\big\{s_t\geq\eps\big\}\Big)\\
\label{eqIHSWP1}
& \leq &
\sum_{t\geq n}\Prob \big(s_t\geq\eps\big)
\ \leq \ \frac{1}{\eps}\sum_{t=n}^\infty\Expect s_t\ \toinfty n\ 0.
\eqa
Actually, (\ref{eq:Solomonoff}) yields an even stronger assertion,
since it characterizes the {\it speed of convergence} by the
quantity on the right hand side. Precisely, the expected number of
times $t$ in which $\xi(a|x\ltt)$ deviates by more than
$\eps$ from $\mu(a|x\ltt)$ is finite and bounded by
$\ln w_\mu^{-1}/\eps^2$, and the probability that the number of
$\eps$-deviations exceeds
$\ln w_\mu^{-1}\over\eps^2\delta$ is smaller than $\delta$.
(However, we \emph{cannot} conclude a convergence rate of
$s_t=o(\frac{1}{t})$ from (\ref{eq:Solomonoff}), since the quadratic differences
generally do not decrease monotonically.)

Since we will encounter this type of convergence
(\ref{eq:Solomonoff}) frequently in the following, we call it {\it
convergence in mean sum (i.m.s)}:
\beq
\label{eq:IMS}
\ph\toims\mu\quad\Longleftrightarrow\quad
\exists\ C>0:\ \sum_{t=1}^\infty \Expect \sum_{a\in\calX}
  \Big( \mu(a|x\ltt)-\ph(a|x\ltt) \Big)^2 \! < \infty.
\eeq
Then Theorem \ref{Theorem:Solomonoff} states that the $\xi$
predictions converge to the $\mu$ predictions i.m.s., or ``$\xi$
converges to $\mu$ i.m.s." for short. By (\ref{eqIHSWP1}),
convergence i.m.s.\ implies almost sure convergence (with respect
to the true distribution $\mu$). Note that in contrast,
convergence in the mean, i.e.\
$\Expect[\sum_a(\mu(a|x\ltt)-\ph(a|x\ltt))^2]\toinfty{t} 0$, only
implies convergence in probability.

\begin{MyPar}{Probabilities vs.\ Description Lengths}
\label{par:prob}
By the Kraft inequality, each (semi)\-measure can be associated
with a code length or {\it complexity} by means of the negative
logarithm, where all (binary) codewords form a prefix-free set.
The converse holds as well. We introduce the abbreviation
\beq
\label{eq:K}
\K\ \ldots=-\lb\ldots, \mbox{ e.g. } \K\nu(x)=-\lb\nu(x)
\eeq
for a semimeasure $\nu$ and $x\in\calX^*$ and $\K\xi(x)=-\lb
\xi(x)$ for the Bayes mixture $\xi$. It is common to ignore
the somewhat irrelevant restriction that code lengths must be
integer. In particular, given a class of semimeasures $\calC$
together with weights, each weight $w_\nu$ corresponds to a
description length or complexity
\beq
\label{eq:Kw}
\K w(\nu)=-\lb w_\nu.
\eeq
It is often only a matter of notational convenience if description
lengths or probabilities are used, but description lengths are
generally preferred in Algorithmic Information Theory. Keeping the
equivalence in mind, we will develop the general theory in terms
of probabilities, but formulate parts of the results in universal
sequence prediction rather in terms of complexities.
\end{MyPar}

%%%%%%%%%%%%%%%%%%%%%%%%%%%%%%%%%%%%%%%%%%%%%%%%%%%%%%%%%%%%%%%
\section{MDL Estimator and Predictions}\label{secMDL}
%%%%%%%%%%%%%%%%%%%%%%%%%%%%%%%%%%%%%%%%%%%%%%%%%%%%%%%%%%%%%%%

Assume that $\calC$ is a countable class of semimeasures
together with weights $(w_\nu)_{\nu\in\calC}$, and
$x\in\calX^*$ is some string. Then the {\it maximizing
element} $\nu^x$, often called MAP (maximum a posteriori)
estimator, is defined as
\beq
\label{eqMaxElem}
  \nu^x = \nu^x_{[\calC]}=\arg\max_{\nu\in\calC}\{w_\nu\nu(x)\}.
\eeq
In case of a tie, we need not specify the further choice at
this point, just pick any of the maximizing elements. But for
concreteness, you may think that ties are broken in favor of
larger prior weights. The maximum is always attained in
(\ref{eqMaxElem}) since for each
$\eps>0$ at most a finite number of elements fulfil
$w_\nu\nu(x)>\eps$. Observe immediately the correspondence in
terms of {\it description lengths} rather than {\it
probabilities}:
\beqn
  \nu^x = \arg\min_{\nu\in\calC}\big\{\K w(\nu)+\K\nu(x)\big\}.
\eeqn
Then the {\it minimum description length principle} is obvious:
$\nu^x$ minimizes the joint description length of the model plus
the data given the model%
\footnote{\label{footnoteMDL}The term MAP estimator is more
precise. For two reasons, our definition might not be
considered as MDL in the strict sense. First, MDL is often
associated with a specific prior, while we admit arbitrary
priors (compare the discussion section at the end of this
paper). Second, when coding some data
$x$, one can exploit the fact that once the distribution $\nu^x$ is
specified, only data which leads to this $\nu^x$ needs to be
considered. This allows for a description shorter than $\K
w(\nu^x)$. Nevertheless, the
\emph{construction principle} is commonly termed MDL, compare
e.g.\ the ``ideal MDL" in \cite{Vitanyi:00}.}
(see (\ref{eq:K}) and (\ref{eq:Kw})).
As explained before, we stick to the product notation.

For notational simplicity, let $\nu^*(x)=\nu^x(x)$. The {\it
two-part MDL estimator} is defined by
\beqn
  \rrho(x) = \rrho_{[\calC]}(x) = w_{\nu^x}\nu^x(x) = \max_{\nu\in\calC}\{w_\nu \nu(x)\}.
\eeqn
So $\rrho$ chooses the maximizing element with respect to its
argument. We may also use the version $\rrho^y(x) :=
w_{\nu^y}\nu^y(x)$ for which the choice depends on the
superscript instead of the argument. Note that the use of the
term ``estimator" is non-standard, since $\rrho$ is a product
of the estimator $\nu^*$ (this use is standard) and its prior
weight. There will be no confusion between these two meanings
of ``estimator" in the following.

For each $x,y\in\calX^*$,
\beq
\label{eq:xirho}
\xi(x)\geq\rrho(x)\geq\rrho^y(x)
\eeq
is immediate. If $\calC$ contains only measures, we have
$\sum_a \rrho(xa) \geq \sum_a \rrho^x(xa) = \rrho^x(x) =
\rrho(x)$
for all $x\in\calX^*$,
so $\rrho$ has some ``anti-semimeasure" property. If $\calC$
contains semimeasures, no semimeasure or anti-semimeasure property
can be established for $\rrho$.

We can define MDL predictors according to
(\ref{eq:basicprediction}). There are basically {\it two} possible
ways to use MDL for prediction.

\begin{Def}
\label{Def:DynamicMDL}
The {\em dynamic} MDL predictor is defined as
\beqn
  \rrho(a|x) = \frac{\rrho(xa)}{\rrho(x)}
  = \frac{\rrho^{xa}(xa)}{\rrho^x(x)}.
\eeqn
That is, we look for a short description of $xa$ and relate it to
a short description of $x=x\ltt$. We call this dynamic since for
each possible $a$ we have to find a new MDL estimator. This is the
closest correspondence to the Bayes mixture $\xi$-predictor.
\end{Def}

\begin{Def}
\label{Def:StaticMDL}
The {\em static} MDL predictor is given by
\beqn
  \rrho^{\mathrm{static}}(a|x)
  = \rrho^x(a|x) = \frac{\rrho^x(xa)}{\rrho(x)}
  = \frac{\rrho^x(xa)}{\rrho^x(x)}
  = \frac{\nu^x(xa)}{\nu^x(x)}.
\eeqn
Here obviously only {\it one} MDL estimator $\rrho^x$ has to be
identified. This is usually more efficient in practice.
\end{Def}

We will define another MDL predictor, the
\emph{hybrid} one, in Section \ref{secHybrid}. It can be
paraphrased as ``do dynamic MDL but drop weights". We will see
that its predictive performance is weaker.

The range of the static MDL predictor is obviously contained in
$[0,1]$. For the dynamic MDL predictor, this holds by
\beq
\label{eq:rhoin01}
\rrho^x(x)\geq\rrho^{xa}(x)\geq\rrho^{xa}(xa).
\eeq

Static MDL is omnipresent in machine learning and
applications, see also Section \ref{secApp}. In fact, many
common prediction algorithms can be abstractly understood as
static MDL, or rather as approximations. Namely, if a
prediction task is accomplished by building a {\it model} such
as a neural network with a suitable
regularization\footnote{There are however regularization
methods which cannot be interpreted in this way but build on a
different theoretical foundation, such as structural risk
minimization.} to prevent ``overfitting", this is just
searching an MDL estimator within a certain class of
distributions. After that, only this model is used for
prediction. Dynamic MDL is applied more rarely due to its
larger computational effort. For example, the similarity
metric proposed in \cite{Li:03} can be interpreted as (a
deterministic variant of) dynamic MDL.

We will need to convert our MDL predictors to {\it measures}
by means of {\it normalization}. If $\ph:\calX^*\to[0,1]$ is
any function, then
\beqn
  \ph\_norm(a|x)
  \ := \ \frac{\ph(a|x)}{\sum_{b\in\calX}\ph(b|x)}
  \ = \ \frac{\ph(xa)}{\sum_{b\in\calX}\ph(xb)}
\eeqn
is a measure (assume that the denominator is different from
zero, which is always true with probability 1 (w.p.1) if $\ph$ is an
MDL predictor). This procedure is known as {\it Solomonoff
normalization} \cite{Solomonoff:78,Li:97} and results in
\beqn
  \ph\_norm(x\leqn) = \frac{\ph(x\leqn)}{\ph(\epstr)}
  \prod_{t=1}^n \frac{\ph(x\ltt)}{\sum_{a\in\calX}\ph(x\ltt a)}
  ={\ph(x\leqn)\over\ph(\epstr)N_\ph(x_{<n})},
\eeqn
where
\beq\label{Eq:Normalizer}
  N_\ph(x) = \prod_{t=1}^{\l(x)+1} \frac{\sum_{a\in\calX}\ph(x\ltt a)}{\ph(x\ltt)}
\eeq
is the normalizer.

We conclude this section with a simple example.

\begin{MyPar}{Bernoulli and i.i.d.\ classes}
\label{ex:iid}
Let $n\in\NNN$, $\calX=\{1,\ldots,n\}$, and
\beqn
  \calC=\Big\{\nu_\th(x_{1:t})=\th_{x_1}\!\cdot...\!\cdot\th_{x_t} : \th\in\Theta\Big\}
  \mbox{\quad with\quad} \Theta=\Big\{\th\in([0,1]\cap\QQQ)^n:\sum_{i=1}^n \th_i=1\Big\}
\eeqn
be the set of all rational probability vectors with any prior
$(w_\th)_{\th\in\Theta}$. Each $\th\in\Theta$ generates sequences $x\ltinf$
of {\it independently identically distributed (i.i.d.)} random
variables such that $\Prob(x_t=i)=\th_i$ for all $t\geq 1$ and
$1\leq i\leq n$. If $x\leqt$ is the initial part of a sequence and
$\alpha\in\Theta$ is defined by $\alpha_i=\frac{1}{t}|\{s\leq t:x_s=i\}|$,
then it is easy to see that
\beqn
\nu^{x\leqt}=\arg\min_{\th\in\Theta}\left\{ \K w(\th)\!\cdot\!\ln 2
+ t\!\cdot\!D(\alpha\|\th)\right\},
\eeqn
where $D(\alpha\|\th):= \sum_{i=1}^n
\alpha_i\ln\frac{\alpha_i}{\th_i}$ is the {\it Kullback-Leibler
divergence}. If $|\calX|=2$, then $\Theta$ is also called a {\it
Bernoulli class}, and one usually takes the binary alphabet
$\calX=\BBB$ in this case. %=\{0,1\}$
\end{MyPar}

%%%%%%%%%%%%%%%%%%%%%%%%%%%%%%%%%%%%%%%%%%%%%%%%%%%%%%%%%%%%%%%
\section{Dynamic MDL}\label{secDynamic}
%%%%%%%%%%%%%%%%%%%%%%%%%%%%%%%%%%%%%%%%%%%%%%%%%%%%%%%%%%%%%%%

We may now develop convergence results, beginning with the dynamic
MDL predictor from Definition \ref{Def:DynamicMDL}. The following
simple lemma is crucial for all subsequent proofs.

\begin{Lemma} \label{Lemma:DiffIsSemimeasure}
For an arbitrary class of (semi)measures $\calC$, we have
\bqan
& (i) &\  \rrho(x)-\sum_{a\in\calX}\rrho(xa)\ \leq\
  \xi(x)-\sum_{a\in\calX}\xi(xa) {\rm\ and}\\
& (ii) &\  \rrho^x(x)-\sum_{a\in\calX}\rrho^x(xa)\ \leq\
  \xi(x)-\sum_{a\in\calX}\xi(xa)
\eqan
for all $x\in\calX^*$. In particular,
$\xi-\rrho$ is a semimeasure.
\end{Lemma}

\begin{Proof}
For all $x\in\calX^*$, with $f:=\xi-\rrho$ we have
\bqan
  \sum_{a\in\calX} f(xa)
  & = & \sum_{a\in\calX} \Big(\xi(xa)-\rrho(xa)\Big)
  \leq \sum_{a\in\calX} \Big(\xi(xa)-\rrho^x(xa)\Big)\\
  & =& \sum_{\nu\in\calM\setminus\{\nu^x\}} \sum_{a\in\calX} w_\nu\nu(xa)
  \leq \sum_{\nu\in\calM\setminus\{\nu^x\}} w_\nu\nu(x)
  = \xi(x)-\rrho(x)
  = f(x).
\eqan
The first inequality follows from $\rrho^x(xa)\leq \rrho(xa)$, and
the second one holds since all $\nu$ are semimeasures. Finally,
$f(x)=\xi(x)-\rrho(x)=\sum_{\nu\in\calM\setminus\{\nu^x\}}
w_\nu\nu(x) \geq 0$ and
$f(\epstr)=\xi(\epstr)-\rrho(\epstr)\leq 1$. Hence $f$ is a
semimeasure.
\end{Proof}

The following proposition demonstrates how simple it can be to
obtain a convergence result, however a weak one. Various
similar results have been already obtained in the past, e.g.\
in \cite{Blackwell:62,Barron:85}.

\begin{Prop} \label{Prop:RhoConvergesMart}
For any class of (semi)measures $\calC$ containing the true
distribution $\mu$, we have
\beqn
  \frac{\rrho(x_t|x\ltt)}{\mu(x_t|x\ltt)}\to 1\quad w.\mu.p.1
\eeqn
\end{Prop}

\begin{Proof}
Since $\xi-\rrho$ is a positive semimeasure by Lemma
\ref{Lemma:DiffIsSemimeasure}, $\frac{\xi-\rrho}{\mu}$ is a
positive super-martingale. By Doob's martingale convergence
theorem (see e.g.\ \cite{Doob:53} or \cite{Chow:88} or any
textbook on advanced probability theory), it therefore converges
on a set of $\mu$-measure one. Moreover, $\frac{\xi}{\mu}$
converges on a set of measure one, being a positive
super-martingale as well \cite[Thm.5.2.2]{Li:97}. Thus
$\frac{\rrho}{\mu}$ must converge on a set of measure one. We
denote this limit by $f$ and observe that $f\geq w_\mu$ since
$\frac{\rrho}{\mu}\geq w_\mu$ everywhere. On this set of measure
one, the denominator $\rrho(x\ltt)/\mu(x\ltt)$ of
\beqn
  \frac{\rrho(x\leqt)/\mu(x\leqt)}{\rrho(x\ltt)/\mu(x\ltt)} =
  \frac{\rrho(x_t|x\ltt)}{\mu(x_t|x\ltt)}
\eeqn
converges to $f>0$, and so does the numerator.
The whole fraction thus converges to one, which was to be shown.
\end{Proof}

Proposition \ref{Prop:RhoConvergesMart} gives only a statement
about ``on-sequence" ($\rrho(x_t|x\ltt)$) convergence of the
$\rrho$-predictions. Indeed, no conclusion about ``off-sequence"
convergence, i.e.\ $\rrho(a|x\ltt)$ for arbitrary $a\in\calX$,
can be drawn from the proposition, not even in the
deterministic case. There, the true measure $\mu$ is
concentrated on the particular sequence $x_{<\infty}$. So for
$a\neq x_t$, we have $\mu(x\ltt a)=0$, and thus no assertion
for $\rrho(a|x\ltt)$ can be made. On the other hand, an
off-sequence result is essential for prediction: Even if on
the \emph{correct} next symbol the predictive probability is
very close to the true value, we must be sure that this is so
also for all \emph{alternatives}. This is particularly
important if we base some decision on the prediction; compare
Section \ref{subsec:Lossbounds}.

The following theorem closes this gap. In addition, it
provides a statement about the speed of convergence. In order
to prove it, we need a lemma establishing a relation between
the square distance and the Kullback-Leibler distance, which
is proven for instance in \cite[Sec.3.9.2]{Hutter:04uaibook}.

\begin{Lemma} \label{Lemma:EntropyIneq}
Let $\mu$ and $\rho$ be measures on $\calX$, then
\beqn
  \sum_{a\in\calX}\big(\mu(a)-\rho(a)\big)^2 \leq
  \sum_{a\in\calX} \mu(a)\ln \frac{\mu(a)}{\rho(a)}.
\eeqn
\end{Lemma}

\begin{Theorem} \label{Theorem:RhoConvergesIMS}
For any class of (semi)measures $\calC$ containing the true
distribution $\mu$ (which is a measure), we have
\beqn
  \sum_{t=1}^\infty \ \Expect \sum_{a\in\calX} \big(\mu(a|x\ltt)-\rrho\_norm(a|x\ltt)\big)^2
  \leq w_\mu^{-1} + \ln w_\mu^{-1}.
\eeqn
That is, $\rrho\_norm(a|x\ltt)\toims\mu(a|x\ltt)$ (see
(\ref{eq:IMS})), which implies
$\rrho\_norm(a|x\ltt)\to\mu(a|x\ltt)$ with $\mu$-probability one.
\end{Theorem}

\begin{Proof}
Let $n\in\NNN$. From Lemma \ref{Lemma:EntropyIneq}, we know
\bqa
  \nonumber
  \lefteqn{\sum_{t=1}^n \ \Expect \sum_{a\in\calX} \big(\mu(a|x\ltt)-\rrho\_norm(a|x\ltt)\big)^2
  \leq \sum_{t=1}^n \ \Expect \sum_{a\in\calX} \mu(a|x\ltt)
  \ln \frac{\mu(a|x\ltt)}{\rrho\_norm(a|x\ltt)}
  }\\
  & = & \sum_{t=1}^n \ \Expect \ln \frac{\mu(x_t|x\ltt)}{\rrho\_norm(x_t|x\ltt)}
  = \sum_{t=1}^n \ \Expect \left[
  \ln\frac{\mu(x_t|x\ltt)}{\rrho(x_t|x\ltt)}
  + \ln \frac{\sum_{a\in\calX}\rrho(x\ltt
  a)}{\rrho(x\ltt)}\right].\quad
  \label{Eq:TheoremRhoConvergesIMS0}
\eqa
Then we can estimate
\beq \label{Eq:TheoremRhoConvergesIMS00}
  \sum_{t=1}^n \ \Expect \ln\frac{\mu(x_t|x\ltt)}{\rrho(x_t|x\ltt)}
  \ = \ \Expect\ \ln\prod_{t=1}^n \frac{\mu(x_t|x\ltt)}{\rrho(x_t|x\ltt)}
  \ = \ \Expect\ \ln\frac{\mu(x\leqn)}{\rrho(x\leqn)}
  \ \leq \ \ln w_\mu^{-1},
\eeq
since always $\frac{\mu}{\rrho}\leq w_\mu^{-1}$. Moreover, by
setting $x=x\ltt$, using $\ln u\leq u-1$, adding an always
positive max-term, and finally using
$\frac{\mu}{\rrho}\leq w_\mu^{-1}$ again, we obtain
\bqa
\nonumber
\lefteqn{
  \Expect \ \ln \frac{\sum_a\rrho(x\ltt a)}{\rrho(x\ltt)}
  \leq \Expect\left[\frac{\sum_a\rrho(xa)}{\rrho(x)}-1\right]
   = \sum_{\zwidths{\l(x)=t-1}}
   \frac{\mu(x)\Big[\big(\sum_a\rrho(xa)\big)-\rrho(x)\Big]}{\rrho(x)}
} \\
\nonumber
  & \leq &
  \sum_{\zwidths{\l(x)=t-1}}
\frac{\mu(x)\Big[\left(\sum_{a\in\calX}\rrho(xa)\right)-\rrho(x)+
  \max\left\{0,\rrho(x)-\sum_{a\in\calX}\rrho(xa)\right\}\Big]}{\rrho(x)}
\\
\label{Eq:TheoremRhoConvergesIMS1}
& \leq &
  w_\mu^{-1}
  \sum_{\l(x)=t-1} \left[\Big(\sum_{a\in\calX}\rrho(xa)\Big)-\rrho(x)+
  \max\Big\{0,\rrho(x)-\sum_{a\in\calX}\rrho(xa)\Big\}\right] .
\eqa
If $\calC$ contains only measures, the max-term is not necessary,
since $\rrho$ is an ``anti-semimeasure'' in this case. We proceed
by observing
\beq \label{Eq:TheoremRhoConvergesIMS2}
  \sum_{t=1}^n \sum_{\l(x)=t-1}\Big[\Big(\sum_{a\in\calX}\rrho(xa)\Big)-\rrho(x)\Big]
  = \sum_{t=1}^n\Big[\sum_{\zwidths{\l(x)=t}}\rrho(x)-
  \sum_{\zwidths{\l(x)=t-1}}\rrho(x)\Big]
  = \Big[\sum_{\zwidths{\l(x)=n}}\rrho(x)\Big]-\rrho(\epstr)
\eeq
which is true since for successive $t$ the positive and negative
terms cancel. From Lemma \ref{Lemma:DiffIsSemimeasure} we know
$  \rrho(x)-\sum_{a\in\calX}\rrho(xa)\ \leq\
  \xi(x)-\sum_{a\in\calX}\xi(xa)$
and therefore
\bqa
  \nonumber
  \sum_{t=1}^n \sum_{\zwidths{\quad \l(x)=t-1}}
   \max\Big\{0,\rrho(x)-\sum_{a\in\calX}\rrho(xa)\Big\}
  & \leq & \sum_{t=1}^n \sum_{\zwidths{\quad \l(x)=t-1}}
  \max\Big\{0,\xi(x)-\sum_{a\in\calX}\xi(xa)\Big\}
  \\
  = \sum_{t=1}^n \sum_{\zwidths{\quad \l(x)=t-1}}\Big[\xi(x)-\sum_{a\in\calX}\xi(xa)\Big]
  & = &  \xi(\epstr)-\sum_{\l(x)=n}\xi(x).
  \label{Eq:TheoremRhoConvergesIMS3}
\eqa
Here we have again used the fact that positive and negative terms
cancel for successive $t$, and moreover the fact that $\xi$ is a
semimeasure. Combining (\ref{Eq:TheoremRhoConvergesIMS1}),
(\ref{Eq:TheoremRhoConvergesIMS2}) and
(\ref{Eq:TheoremRhoConvergesIMS3}), and observing
$\rrho\leq\xi\leq 1$, we obtain
\beq \label{Eq:TheoremRhoConvergesIMSNormalizer}
  \sum_{t=1}^n \ \Expect \ln \frac{\sum_a\rrho(x\ltt a)}{\rrho(x\ltt)}
\leq w_\mu^{-1}
   \left[\xi(\epstr)-\rrho(\epstr)+\sum_{\zwidths{\l(x)=n}}\big(\rrho(x)-\xi(x)\big)\right]
  \leq w_\mu^{-1}\xi(\epstr) \leq w_\mu^{-1}.
\eeq
Therefore, (\ref{Eq:TheoremRhoConvergesIMS0}), (\ref{Eq:TheoremRhoConvergesIMS00})
and (\ref{Eq:TheoremRhoConvergesIMSNormalizer}) finally prove the assertion.
\end{Proof}

We point out again that the proof gets a bit simpler if
$\calC$ contains only measures, since then
(\ref{Eq:TheoremRhoConvergesIMS3}) becomes irrelevant.
However, this case doesn't give a tighter bound.

This is the first convergence result ``in mean sum", see
(\ref{eq:IMS}). It implies both on-sequence and off-sequence
convergence. Moreover, it asserts the convergence is ``fast" in
the sense that the sum of the total expected deviations is bounded
by $w_\mu^{-1}+\ln w_\mu^{-1}$. Of course, $w_\mu^{-1}$ can be
very large, namely
$w_\mu^{-1}=2^{\K w(\mu)}$.
The following example will show that this bound is sharp (save for
a constant factor). Observe that in the corresponding result for
mixtures, Theorem \ref{Theorem:Solomonoff}, the bound is much
smaller, namely
$\ln w_\mu^{-1}=\K w(\mu)\ln 2$.

\begin{Example}\label{Ex:LowerBound}
Let $\calX=\{0,1\}$, $N\geq 1$ and
$\calC=\{\nu_1,\ldots,\nu_{N-1},\mu\}$. Each $\nu_i$ is a
deterministic measure concentrated on the sequence
$z\ltinf^{(i)}=1^{i-1}0^\infty$, while the true distribution $\mu$
is deterministic and concentrated on $x\ltinf=1^\infty$. Let
$w_{\nu_i}=w_\mu=\frac{1}{N}$ for all $i$. Then $\mu$ generates
$x\ltinf$, and for each $t\leq N-1$ we have
$\rrho\_norm(0|x\ltt)=\rrho\_norm(1|x\ltt)=\frac{1}{2}$. Hence,
$\sum_t\Expect\sum_a\big(\mu(a|x\ltt)-\rrho\_norm(a|x\ltt\big))^2=\frac{1}{2}(N-1)\eqm
w_\mu^{-1}$.
In Example \ref{ex:bernoulli} we will even see a case where
the model class contains only Bernoulli distributions and
nevertheless the exponential bound is sharp.
\end{Example}

The next result implies that convergence holds also for the
un-normalized dynamic MDL predictor.

\begin{Theorem} \label{thNormalizer} \label{thRhoNorm}
For any class of (semi)measures $\calC$ containing the true
distribution $\mu$, we have
\bqan
  & (i) &
  \sum_{t=1}^\infty \Expect \left|\,
  \ln \sum_{a\in\calX}\rrho(a|x\ltt) \right|
  \ \leq \ 2 w_\mu^{-1} {\rm\quad and}
\\
  & (ii) & \sum_{t=1}^\infty \Expect \sum_{a\in\calX}
  \Big| \rrho\_norm(a|x\ltt)-\rrho(a|x\ltt) \Big|
  \ = \
  \sum_{t=1}^\infty \Expect
  \Big| 1- \sum_{a\in\calX}\rrho(a|x\ltt) \Big|
  \ \leq \ 2 w_\mu^{-1}.
\eqan
\end{Theorem}

\begin{Proof}
$(i)$ Define $u^+=\max\{0,u\}$ for $u\in\RRR$, then for $x:=x\ltt\in\calX^{t-1}$ we have
\bqan
  \lefteqn{
  \Expect \Big|\,\ln \sum_{a\in\calX}\rrho(a|x) \Big| =
  \Expect \left|\,
  \ln {\sum_a\rrho(xa)\over\rrho(x)} \right|
  = \Expect \left[ \Big(\ln{\sum_a\rrho(xa)\over\rrho(x)}\Big)^+
  +\Big(\ln{\rrho(x)\over\sum_a\rrho(xa)}\Big)^+\right]
}\\
  & \leq & \Expect {\big(\sum_a\rrho(xa)-\rrho(x)\big)^+\over\rrho(x)}
  \ + \ \Expect {\big(\rrho(x)-\sum_a\rrho(xa)\big)^+\over\sum_a\rrho(xa)}
\\
  & = & \sum_{\zwidths{\l(x)=t-1}} {\mu(x)\big(\sum_a\rrho(xa)-\rrho(x)\big)^+\over\rrho(x)}
  \ + \sum_{\zwidths{\l(x)=t-1}} {\mu(x)\big(\rrho(x)-\sum_a\rrho(xa)\big)^+\over\sum_a\rrho(xa)}
\\
  & \leq & w_\mu^{-1} \sum_{\zwidths{\l(x)=t-1}}
  \big({\textstyle\sum_a\rrho(xa)}-\rrho(x)\big)^+
  \ + \ w_\mu^{-1} \sum_{\zwidths{\l(x)=t-1}}
  \big(\rrho(x)-{\textstyle\sum_a\rrho(xa)}\big)^+
\\
  & = & w_\mu^{-1} \sum_{\zwidths{\l(x)=t-1}} |\rrho(x)\!-\!{\textstyle\sum_a\rrho(xa)}|
  = w_\mu^{-1} \sum_{\zwidths{\l(x)=t-1}} \left[\textstyle\sum_a\rrho(xa)-\rrho(x)
  + 2\big(\rrho(x)\!-\!{\textstyle\sum_a\rrho(xa)}\big)^+\right]
\eqan
Here, $|u|=u^++(-u)^+=-u+2u^+$, $\ln u\leq u-1$, and
$\rrho\geq w_\mu\mu$ have been used, the latter implies also
$\sum_a\rrho(xa)\geq w_\mu\sum_a\mu(xa)=w_\mu\mu(x)$.
The last expression in this (in)equality chain, when summed over
$t=1...\infty$ is bounded by $2w_\mu^{-1}$ by essentially the same
arguments (\ref{Eq:TheoremRhoConvergesIMS1}) -
(\ref{Eq:TheoremRhoConvergesIMSNormalizer}) as in the proof of
Theorem \ref{Theorem:RhoConvergesIMS}.

$(ii)$ Let again $x:=x\ltt$ and use
$\rrho\_norm(a|x)=\rrho(a|x)/\sum_b\rrho(b|x)$ to obtain
\bqa\label{pr:rnr}
  \!\!\!\!\!\sum_a \Big| \rrho\_norm(a|x)-\rrho(a|x) \Big|
  & = & \sum_a {\rrho(a|x)\over\sum_b\rrho(b|x)}\Big| 1\!-\!\sum_b\rrho(b|x) \Big|
   =  \Big| 1\!-\!\sum_b\rrho(b|x) \Big|\\ \nonumber
  & = & {\big(\sum_a\rrho(xa)-\rrho(x)\big)^+\over\rrho(x)}
   +  {\big(\rrho(x)-\sum_a\rrho(xa)\big)^+\over\rrho(x)}.
\eqa
Then take the expectation $\Expect$ and the sum
$\sum_{t=1}^\infty$ and proceed as in $(i)$.
\end{Proof}

\begin{Cor}
\label{Cor:MuRho}
For any class of (semi)measures $\calC$ containing the true
distribution $\mu$, we have
\beqn
\sum_{t=1}^\infty \Expect \sum_{a\in\calX}
  \Big( \mu(a|x\ltt)-\rrho(a|x\ltt) \Big)^2 \ \leq \ 8 w_\mu^{-1}.
\eeqn
That is, $\rrho(a|x\ltt)\toims\mu(a|x\ltt)$ (see (\ref{eq:IMS})).
\end{Cor}

\begin{Proof}
For two functions $\ph_1,\ph_2:\calX^*\to [0,1]$, let
\beq
\label{eq:Delta}
\Delta(\ph_1,\ph_2)=\left(\sum_{t=1}^\infty \Expect \sum_{a\in\calX}\Big(
\ph_1(a|x\ltt)-\ph_2(a|x\ltt)\Big)^2\right)^{\frac{1}{2}}.
\eeq
Then the triangle inequality holds for $\Delta(\cdot,\cdot)$,
since $\Delta$ is (proportional to) an Euclidian distance
(2-norm). Moreover,
$\Delta(\mu,\rrho\_norm)\leq\sqrt{2 w_\mu^{-1}}$ by Theorem
\ref{Theorem:RhoConvergesIMS} and
$\ln w_\mu^{-1}\leq w_\mu^{-1}-1\leq w_\mu^{-1}$.
We also have $\Delta(\rrho\_norm,\rrho)\leq\sqrt{2 w_\mu^{-1}}$ by
multiplying $|\rrho\_norm-\rrho|$ in Theorem \ref{thRhoNorm}$(ii)$
with another $|\rrho\_norm-\rrho|$. Note
$|\rrho\_norm-\rrho|\leq 1$, since both
$\rrho(a|x),\rrho\_norm(a|x)\in[0,1]$,  for $\rrho$ this holds by
(\ref{eq:rhoin01}). This implies
$\Delta(\mu,\rrho)\leq
\Delta(\mu,\rrho\_norm)+\Delta(\rrho\_norm,\rrho)\leq 2\sqrt{2
w_\mu^{-1}}$.
\end{Proof}

\begin{Cor}
For almost all $x_{<\infty}\in\calX^\infty$, the normalizer
$N_\rrho$ defined in (\ref{Eq:Normalizer}) converges to a number
which is finite and greater than zero, i.e.\
$0<N_\rrho(x_{<\infty})<\infty$. Moreover, the sum of the MDL
posterior estimates converges to one almost surely,
\beq
  \label{Eq:TheoremRhoNormalizerConverges}
  \sum_{a\in\calX} \rrho(a|x\ltt) = \frac{\sum_{a\in\calX}\rrho(x\ltt a)}{\rrho(x\ltt)} \to 1
  {\rm\quad as\quad} t\to\infty \quad w.\mu.p.1.
\eeq
\end{Cor}

\begin{Proof}
Theorem \ref{thNormalizer} implies that with probability one, the sum
$ \sum_1^n \big| \ln \frac{\sum_a\rrho(x\ltt a)}{\rrho(x\ltt)} \big| $
is bounded in $n$, hence converges absolutely, hence also the limit
\beqn
  \ln N_\rrho(x_{<\infty})
  \ = \
  \sum_{t=1}^\infty \ln \frac{\sum_{a\in\calX}\rrho(x\ltt a)}{\rrho(x\ltt)}
\eeqn
exists and is finite. For these sequences,
$0<N_\rrho(x_{<\infty})<\infty$ and
(\ref{Eq:TheoremRhoNormalizerConverges}) follows.
\end{Proof}

%%%%%%%%%%%%%%%%%%%%%%%%%%%%%%%%%%%%%%%%%%%%%%%%%%%%%%%%%%%%%%%
\section{Static MDL}\label{secStatic}
%%%%%%%%%%%%%%%%%%%%%%%%%%%%%%%%%%%%%%%%%%%%%%%%%%%%%%%%%%%%%%%

Static MDL as introduced in Definition \ref{Def:StaticMDL} is
usually more efficient and thus preferred in practice, since
only one MDL estimator has to be computed. The following
technical result will allow to conclude that the static MDL
predictions converge in mean sum like the dynamic ones.

\begin{Theorem} \label{thSMDLBound}
For any class of (semi)measures $\calC$ containing the true
distribution $\mu$, we have
\beqn
  \sum_{t=1}^\infty \Expect
  \sum_{a\in\calX} \Big| \rrho\_norm^{x\ltt}(a|x\ltt) - \rrho^{x\ltt}(a|x\ltt) \Big|
  \ = \ \sum_{t=1}^\infty \Expect
  \Big| 1- \sum_{a\in\calX}\rrho^{x\ltt}(a|x\ltt) \Big|
  \ \leq \ w_\mu^{-1}.
\eeqn
\end{Theorem}

\begin{Proof}
We proceed in a similar way as in the proof of Theorem
\ref{Theorem:RhoConvergesIMS},
(\ref{Eq:TheoremRhoConvergesIMS1}) - (\ref{Eq:TheoremRhoConvergesIMS3}).
From Lemma \ref{Lemma:DiffIsSemimeasure}, we know
$  \rrho(x)-\sum_a\rrho^x(xa)\ \leq\  \xi(x)-\sum_a\xi(xa)$.
Then
\bqan
  \sum_{t=1}^n \Expect
  \Big| 1- \sum_{a\in\calX}\rrho^{x\ltt}(a|x\ltt) \Big|
  & = & \sum_{t=1}^n\ \Expect \frac{\rrho(x\ltt)-\sum_{a\in\calX}\rrho^{x\ltt}(x\ltt a)}{\rrho(x\ltt)}
\\
  & = & \sum_{t=1}^n \sum_{\l(x)=t-1} \mu(x)
  \frac{\rrho(x)-\sum_{a\in\calX}\rrho^x(xa)}{\rrho(x)}
\\
  & \leq & w_\mu^{-1}\ \sum_{t=1}^n \sum_{\l(x)=t-1}
  \left[\rrho(x)-\sum_{a\in\calX}\rrho^x(xa)\right]
\\
  & \leq & w_\mu^{-1}\ \sum_{t=1}^n \sum_{\l(x)=t-1}
  \left[\xi(x)-\sum_{a\in\calX}\xi(xa)\right]
\\
  & \leq & w_\mu^{-1}\left[\xi(\epstr)-\sum_{\l(x)=n}\xi(x)\right]
  \ \leq \ w_\mu^{-1}
\eqan
for all $n\in\NNN$. This implies the assertion. Again we have used
$\frac{\mu}{\rrho}\leq w_\mu^{-1}$ and the fact that positive and
negative terms cancel for successive $t$.
\end{Proof}

\begin{Cor}
\label{Cor:MuRhoStatic}
For any class of (semi)measures $\calC$ containing the true
distribution $\mu$, we have
\bqan
\sum_{t=1}^\infty \Expect \sum_{a\in\calX}
  \Big( \mu(a|x\ltt)-\rrho^{x\ltt}(a|x\ltt) \Big)^2 & \leq & 21 w_\mu^{-1}\und\\
\sum_{t=1}^\infty \Expect \sum_{a\in\calX}
  \Big( \mu(a|x\ltt)-\rrho^{x\ltt}\_norm(a|x\ltt) \Big)^2 & \leq & 32 w_\mu^{-1}.
\eqan
That is, $\rrho^{x\ltt}(a|x\ltt)\toims\mu(a|x\ltt)$ and
$\rrho^{x\ltt}\_norm(a|x\ltt)\toims\mu(a|x\ltt)$.
\end{Cor}

\begin{Proof}
Using $\rrho(xa)\geq\rrho^x(xa)$ and the triangle inequality, we see
\beqn
 \sum_a\Big| \rrho(a|x)-\rrho^x(a|x) \Big|
 = \Big| \sum_a\rrho(a|x)- \sum_a\rrho^x(a|x) \Big|
 \leq \Big| \sum_a\rrho(a|x)- 1 \Big|
      + \Big| 1 - \!\sum_a\rrho^x(a|x) \Big|
\eeqn
With $\Delta(\cdot,\cdot)$ as in (\ref{eq:Delta}), using
$|\rrho-\rrho^x|\leq 1$ we therefore have
\beqn
\Delta^2(\rrho,\rrho^{\mathrm{static}})\leq
\sum_{t=1}^\infty\Expect\sum_a\Big| \rrho(a|x)-\rrho^x(a|x)\Big|
\leq 3w_\mu^{-1}
\eeqn
according to Theorem \ref{thRhoNorm} $(ii)$ and Theorem \ref{thSMDLBound}.
Since $\Delta(\mu,\rrho)\leq 2\sqrt{2 w_\mu^{-1}}$
holds by Corollary \ref{Cor:MuRho}, we obtain
$\Delta(\mu,\rrho^{\mathrm{static}})\leq
\Delta(\mu,\rrho)+\Delta(\rrho,\rrho^{\mathrm{static}})\leq
\sqrt{21 w_\mu^{-1}}$.
Theorem \ref{thSMDLBound} also asserts
$\Delta(\rrho^{\mathrm{static}},\rrho^{\mathrm{static}}\_norm)\leq\sqrt{w_\mu^{-1}}$, hence
$\Delta(\mu,\rrho^{\mathrm{static}}\_norm)\leq \sqrt{32 w_\mu^{-1}}$ follows.
\end{Proof}

\begin{MyPar}{Distance measures}
The total expected square error is not the only possible
choice for measuring distance of distributions and speed of
convergence. In fact, looking at the proof of Theorem
\ref{Theorem:RhoConvergesIMS}, the expected Kullback-Leibler
distance may seem more natural at a first glance. However this
quantity behaves well only under dynamic MDL, not static MDL.
To see this, let
$\calC\cong\{0,\frac{1}{2}\}$ contain two Bernoulli
distributions, both with prior weight $\frac{1}{2}$, and let
$\mu\cong\frac{1}{2}$ be the uniform measure. If the first
symbol happens to be 0, which occurs with probability
$\frac{1}{2}$, then the static MDL estimate is $\nu^0\cong 0$.
Then $D(\mu\|\nu^0)=\infty$, hence the expectation is
$\infty$, too. The quadratic distance behaves locally like the
Kullback-Leibler distance (Lemma \ref{Lemma:EntropyIneq}), but
otherwise is bounded and thus more convenient.
\end{MyPar}

Another possible choice is the \emph{Hellinger distance}
\bqa
\label{eq:Hellinger}
h_t(\mu,\ph)|_{x\ltt}&=&\sum_{a\in\calX}\Big(\sqrt{\mu(a|x\ltt)}-\sqrt{\ph(a|x\ltt)}\Big)^2
\und \\
\label{eq:CumHellinger}
H\leqn(\mu,\ph)&=&\sum_{t=1}^n\Expect h_t(\mu,\ph).
\eqa
Like the square distance, the Hellinger distance is bounded by
both the relative entropy and the absolute distance:
\bqa
\label{eqHellingerKL}
h_t(\mu,\ph) & \leq & \sum_{a\in\calX} \mu(a|x\ltt)\ln\frac{\mu(a|x\ltt)}{\ph(a|x\ltt)} \und\\
\label{eqHellingerAbs}
h_t(\mu,\ph) & \leq & \sum_{a\in\calX}
\Big| \mu(a|x\ltt)- \ph(a|x\ltt) \Big|.
\eqa
The former is e.g.\ shown in \cite[Lem.3.11, p.114]{Hutter:04uaibook},
the latter follows from
$(\sqrt u-\sqrt v)^2\leq|u-v|$ for any $u,v\in\RRR$.
Therefore, the same bounds we have proven for the square
distance also hold for the Hellinger distance; they are
subsumed in Corollary \ref{Cor:MuRhoAll} below. Although for
simplicity of notation we have preferred the square distance
over the Hellinger distance in the presentation so far, in
Sections \ref{subsec:Lossbounds} and \ref{subsec:Regression}
we will meet situations where the quadratic distance is not
sufficient. Then the Hellinger distance will be useful.

The following corollary recapitulates our results and states
convergence i.m.s (and therefore also w.$\mu$-p.1) for all
combinations of un-normalized/normalized and dynamic/static
MDL predictions.

\begin{Cor}
\label{Cor:MuRhoAll}
Let $\calC$ contain the true distribution $\mu$, then%
\bqan
S\ltinf(\mu,\rrho\_norm)\leq 2 w_\mu^{-1}, &&
H\ltinf(\mu,\rrho\_norm)\leq 2 w_\mu^{-1}, \\
S\ltinf(\mu,\rrho)\leq 8 w_\mu^{-1}, &&
H\ltinf(\mu,\rrho)\leq 8 w_\mu^{-1}, \\
S\ltinf(\mu,\rrho^{\mathrm{static}})\leq 21w_\mu^{-1}, &&
H\ltinf(\mu,\rrho^{\mathrm{static}})\leq 21w_\mu^{-1}, \\
S\ltinf(\mu,\rrho^{\mathrm{static}}\_norm)\leq 32w_\mu^{-1},
&& H\ltinf(\mu,\rrho^{\mathrm{static}}\_norm)\leq
32w_\mu^{-1},
\eqan
where $S\ltinf(\mu,\ph)=\sum_t \Expect \sum_a  \big(
\mu(a|x\ltt)-\ph(a|x\ltt) \big)^2$ and $H\ltinf$ is as in (\ref{eq:CumHellinger}).
\end{Cor}

The following example shows that the exponential bound is
sharp (except for a multiplicative constant), even if the
model class contains only Bernoulli distributions. It is
stated in terms of static MDL, however it equally holds for
dynamic MDL.

\begin{Example}\label{ex:bernoulli}
Let $N\geq 1$ and
$\calC\cong\Theta=\{\frac{1}{2}\}\cup\{\frac{1}{2}+2^{-k-1}:1\leq k\leq
N\}$ be a Bernoulli class as discussed at the end of Section
\ref{secMDL}. Let $\mu$ be Bernoulli with parameter
$\frac{1}{2}$, i.e.\ the distribution generating fair coin flips.
Assume that all weights are equally $\frac{1}{N+1}$. Then it is
shown in \cite[Prop. 5]{Poland:04mdlspeed} that
\beqn
  \sum_{t=1}^\infty \Expect \big({\textstyle \frac{1}{2}}-\rrho^{x\ltt}(1|x\ltt)\big)^2\geq \mbox{$\frac{1}{84}$}\big(N-4\big).
\eeqn
So the bound equals $w_\mu^{-1}$ within a multiplicative constant.
\end{Example}

This shows that in general there is no hope to improve the
bounds, even for very simple model classes. But the situation
is not as bad as it might seem. First, the bounds may be
exponentially smaller under certain regularity conditions on
the class and the weights, as \cite{Rissanen:96} and the
positive assertions in \cite{Poland:04mdlspeed} show. It is
open to define such conditions for more general model classes.
Second, the example just given behaves differently than
Example \ref{Ex:LowerBound}. There, the error remains at a
significant level for $O(w_\mu^{-1})$ time steps, which must
be regarded critical. Here in contrast, the error drops to
zero as $\frac{1}{n}$ for a very long time, namely
$O(2^{w_\mu^{-1}})$ steps, and decreases more rapidly only
afterwards. This behavior is tolerable in practice. Recently,
\cite{Li:99,Zhang:04} have proven that this favorable case always
occurs for i.i.d., if the weights satisfy the \emph{light tails}
condition $\sum w_\nu^\alpha\leq 1$ for some $\alpha<1$
\cite{Barron:91}. Precisely, they give a rapidly decaying
bound on the instantaneous error. It is open if similar
results also hold in more general setups than i.i.d. Example
\ref{Ex:LowerBound} shows that at least some additional
assumption is necessary.

%%%%%%%%%%%%%%%%%%%%%%%%%%%%%%%%%%%%%%%%%%%%%%%%%%%%%%%%%%%%%%%
\section{Hybrid MDL}\label{secHybrid}
%%%%%%%%%%%%%%%%%%%%%%%%%%%%%%%%%%%%%%%%%%%%%%%%%%%%%%%%%%%%%%%

So far, we have not cared about what happens if two or more
(semi)measures obtain the same value $w_\nu
\nu(x)$ for some string $x$. In fact, for the previous results,
the {\it tie-breaking strategy} can be completely arbitrary. This
need not be so for all thinkable prediction methods, as we will
see with the hybrid MDL predictor in the subsequent example.

\begin{Def}
\label{Def:HybridMDL}
The {\em hybrid} MDL predictor is given by
$$
  \rrho^{\mathrm{hyb}}(a|x)
  = \frac{\nu^*(xa)}{\nu^*(x)}
$$
(compare (\ref{eqMaxElem})). This can be paraphrased as ``do
dynamic MDL and drop the weights". It is somewhat in-between
static and dynamic MDL.
\end{Def}

\begin{Example}\label{ex:first} Let $\calX=\BBB$ and $\calC$ contain only two
measures, the uniform measure $\lambda$ which is defined by
$\lambda(x)=2^{-\l(x)}$, and another measure
$\nu$ having $\nu(1x)=2^{-\l(x)}$ and $\nu(0x)=0$. The respective
weights are $w_\lambda=\frac{2}{3}$ and $w_\nu=\frac{1}{3}$. Then,
for each $x$ starting with $1$, we have
$w_\nu\nu(x)=w_\lambda\lambda(x)=\frac{1}{3}2^{-\l(x)+1}$. Therefore,
for all $x\ltinf$ starting with $1$ (a set which has uniform measure $\frac{1}{2}$),
we have a tie. If the maximizing element $\nu^*$ is chosen to be $\lambda$ for
even $t$ and $\nu$ for odd $t$, then both static and dynamic MDL
predict probabilities of constantly
\beqn
  \tfrac{1}{2}=\lambda(a|x\ltt)=\nu(a|x\ltt)=
  \frac{w_\lambda\lambda(x\ltt a)}{w_\nu\nu(x\ltt)}=
  \frac{w_\nu\nu(x\ltt a)}{w_\lambda\lambda(x\ltt)}
\eeqn
for all $a\in\BBB$. However, the hybrid MDL predictor values
$\frac{\nu^*(x\ltt a)}{\nu^*(x\ltt)}$ oscillate between ${1\over
4}$ and $1$.
\end{Example}

If the ambiguity in the tie-breaking process is removed, e.g.\
in favor of larger weights, then the hybrid MDL predictor does
converge for this example. We replace (\ref{eqMaxElem}) by
this rule:
\beqn
  \nu^x = \arg\max\big\{w_\nu:\nu\in\{\nu=\arg\max_{\nu\in\calC}w_\nu\nu(x)\}\big\}.
\eeqn
Then, do the hybrid MDL predictions always converge? This is
equivalent to asking if the process of selecting a maximizing
element eventually {\it stabilizes}. If stabilization does not
occur, then hybrid MDL will necessarily fail as soon as the
weights are not equal. A possible counterexample could consist
of two measures the fraction of which oscillates perpetually
around a certain value. We show that this can indeed happen,
even for different reasons.

\begin{Example} \label{Example:NotStabilizeApprox}
Let $\calX$ be binary, $\mu(x)=\prod_{i=1}^{\l(x)}\mu_i(x_i)$ and
$\nu(x)=\prod_{i=1}^{\l(x)}\nu_i(x_i)$ with
\beqn
\mu_i(1)=1-2^{-2\left\lceil\frac{i}{2}\right\rceil} {\rm\ and\ }
\nu_i(1)=1-2^{-2\left\lceil\frac{i+1}{2}\right\rceil+1}.
\eeqn
Then one can easily see that $\mu(111\ldots)=\prod_1^\infty\mu_i(1)>0$,
$\nu(111\ldots)=\prod_1^\infty\nu_i(1)>0$,
and $\frac{\nu(111\ldots)}{\mu(111\ldots)}$ converges and oscillates.
In fact, each sequence having positive measure under $\mu$ and $\nu$ contains
eventually only ones, and the quotient oscillates.
\end{Example}

\begin{figure}[t!] \begin{center}
\scalebox{0.7}{
\includegraphics{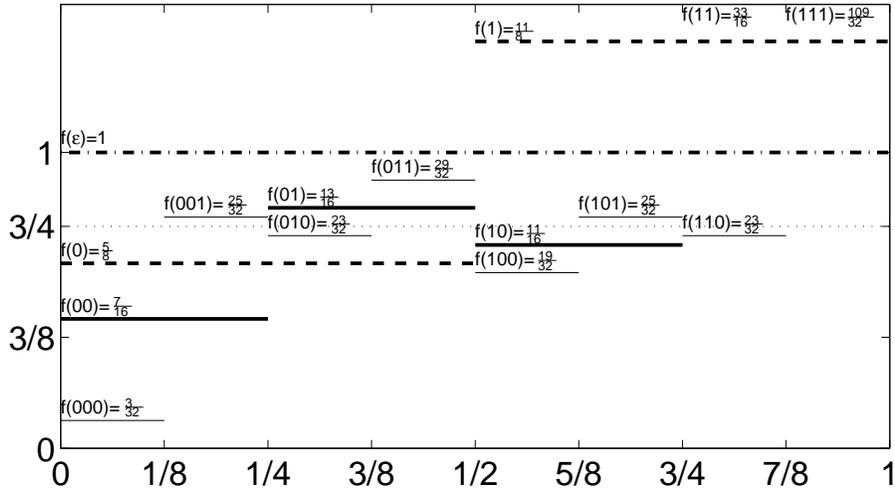}}
\caption{Construction of a martingale that with high probability
converges to $\frac{3}{4}$ oscillating infinitely often.}
\label{Fig:MartingaleEx}
\end{center} \end{figure}

\begin{Example} \label{Example:NotStabilizeDependent}
This example is a little more complex.
We assume the uniform distribution $\lambda$ to be the true
distribution. We now construct a positive martingale $f(\cdot)$ that converges
to $\frac{3}{4}$ with high probability and thereby oscillates
infinitely often.

The martingale is defined on strings $x$ of successively increasing
length. Of course, $f(\epstr):=1$. If $f(x)$ is already defined for strings
of length $n-1$, we extend the definition on strings of length $n$ in
the following way: If $f(x)>\frac{3}{4}$, we set
\bqan
f(x0) & := & \frac{3}{4}-2^{-n-2} {\rm\quad and}\\
f(x1) & := & 2f(x)-\Big(\frac{3}{4}-2^{-n-2}\Big).
\eqan
This guarantees the martingale property
$f(x)=\frac{1}{2}\big(f(x0)+f(x1)\big)$. If $f(x)\leq\frac{3}{4}$
and $f(x)\geq\frac{3}{8}+2^{-n-3}$, then we can similarly define
\bqan
f(x0) & := & 2f(x)-\Big(\frac{3}{4}+2^{-n-2}\Big) {\rm\quad and}\\
f(x1) & := & \frac{3}{4}+2^{-n-2}.
\eqan
However, if $f(x)<\frac{3}{8}+2^{-n-3}$, we cannot proceed in this way,
since $f$ must be positive. Therefore, we set
$ f(x0):=f(x1):=f(x)$ in this case and call those $x$
``dead" strings. Strings that are not dead will be
called ``alive".
A few steps of the construction are shown in Figure \ref{Fig:MartingaleEx}.
For example, it can be observed that the string 000 is dead, all other strings
in the figure are alive.
\end{Example}

It is obvious from the construction that $f(x\leqt)$ is a martingale,
it oscillates and
converges to $\frac{3}{4}$ as $t\to\infty$ for all sequences $x\ltinf$
that always stay alive. The only thing we must show is that many sequences
in fact stay alive.

\begin{Claim}
We have
$\lambda\big(\{x\ltinf:\ \exists t \such_that x\leqt{\rm\ is\ dead}\}\big)\leq\frac{1}{4}$.
\end{Claim}

\begin{Proof}
After the $n$th step, i.e.\ when $f$ has been defined for strings of
length $n$, $f(x)$ assumes the value
\beqn
  a_0^n=\frac{3}{4}-2^{-n-2}
\eeqn
on a set of measure
at most $\frac{1}{2}$. In the next step $n+1$, $f$ is defined to
\beqn
  a_1^n=\frac{3}{4}-2^{-n-1}\Big(1+\frac{1}{4}\Big)
\eeqn
on half of the extended strings. Generally, in the $k$th next step,
$f$ is defined to
\beqn
  a_k^n=\frac{3}{4}-2^{-n+k-2}\Big(\sum_{j=0}^k2^{-2j}\Big)
\eeqn
on a $2^{-k}$ fraction of the extended strings.

The extended strings stay alive as long as $a_k^n\geq\frac{3}{8}+2^{-n-k-3}$
holds. Some elementary calculations show that this is equivalent
to $k\leq n$. So precisely after $n+1$ additional steps, a fraction of $2^{-n-1}$ of
the extended strings die.

We already noted that for $A_n=\{x:\l(x)=n\wedge f(x)=a_0^n\}$,
we have $\lambda(A_n)\leq\frac{1}{2}$.
Thus, $$\lambda\big(\{x\ltinf:x\leqn\in A_n{\rm\ and\ }
x_{1:2n+1}{\rm\ is\ dead}\}\big)\leq 2^{-n-2}.$$
Hence, one can conclude
$$\lambda\big(\{x\ltinf:\ \exists t \such_that x\leqt{\rm\ is\ dead}\}\big)
\leq\sum_{n=1}^\infty 2^{-n-2}=\frac{1}{4},$$
which proves the claim.
\end{Proof}

We now define the measure $\nu$ by
$$\nu(x)=f(x)\cdot\lambda(x)=f(x)\cdot2^{-\l(x)},$$
and set the weights to $w_\lambda=\frac{3}{7}$ and
$w_\nu=\frac{4}{7}$. Then this
provides an example where the maximizing element never stops
oscillating with probability at least $\frac{3}{4}$.

Both examples point out different possible reasons for failure
of stabilizing. Example \ref{Example:NotStabilizeApprox}
works since the measure $\mu$ and
$\nu$ are asymptotically very similar and close to deterministic.
In contrast, in Example \ref{Example:NotStabilizeDependent}
stabilizing fails because of lack of independence:
The quantity $\nu(a|x)$ strongly depends on $x$.
In particular, one can note that even Markovian dependence
may spoil the stabilization, since $\nu(a|x)$ only depends
on the last symbol of $x$.

%%%%%%%%%%%%%%%%%%%%%%%%%%%%%%%%%%%%%%%%%%%%%%%%%%%%%%%%%%%%%%%
\section{Stabilization}\label{secStabilization}
%%%%%%%%%%%%%%%%%%%%%%%%%%%%%%%%%%%%%%%%%%%%%%%%%%%%%%%%%%%%%%%

In the light of the previous section, it is therefore natural
to ask when the maximizing element stabilizes (almost surely).
Barron \cite{Barron:85,Barron:98} has shown that this happens
if all distributions in $\calC$ are \emph{asymptotically
mutually singular}. Under this condition, the true
distribution is even eventually identified almost
surely.\footnote{In general, stabilization does not imply that
the true distribution is identified. Consider for instance a
model class containing two measures: the true measure is
concentrated on $0^\infty$ and has prior weight
$\tfrac{1}{8}$, the other one assigns probability
$\nu(x_t=1)=2^{-t}$ independently of the past
$x\ltt$. Then the maximizing element will remain the incorrect
distribution $\nu$, however with predictions rapidly
converging to the truth.}

The condition of asymptotic mutual singularity holds in many
important cases, e.g.\ if the distributions are i.i.d.
However, one cannot always build on it.\footnote{%
Here is a simple example: let the true measure be
Bernoulli($\frac{1}{2}$) and another measure be a product of
Bernoullis with parameter rapidly converging to $\frac{1}{2}$.
These distributions are not asymptotically mutually singular,
nevertheless a.s. stabilization holds, as we will see.}
Therefore, in this section we give a different approach: In
order to prevent stabilization, it is necessary that the ratio
of two predictive distributions oscillates around the inverse
ratio of the respective weights. Therefore, stabilization must
occur almost surely if the ratio of two predictive
distributions converges almost surely but is not
\emph{concentrated} in the limit. This is satisfied under
appropriate conditions, as we will prove. We start with a
general theorem which allows to conclude almost sure
stabilization in a countable model class, if for any
\emph{pair} of models we have almost sure stabilization.

\begin{Theorem} \label{Theorem:MDLstabilizesF2C}
Let $\calC$ be a countable class of (semi)measures containing
the true measure $\mu$. Assume that for each two
$\nu_1,\nu_2\in\calC$ the maximizing element chosen from
$\{\nu_1,\nu_2\}$ eventually stabilizes almost surely. Then
also the maximizing element chosen from all of $\calC$
stabilizes almost surely.
\end{Theorem}

\begin{Proof} It is immediate that the maximizing element
chosen from any finite subset of $\calC$ stabilizes almost
surely. Now, for all $\nu\in\calC$ and $c>0$, define the set
$A_c^\nu$ by
\beqn
A_c^\nu = \left\{x\ltinf:\exists\ t\geq
1\such_that\frac{\nu(x\leqt)}{\mu(x\leqt)}\geq c\right\}.
\eeqn
Then we have
\bqan
\mu(A_c^\nu) & = & \mu\left(\bigcup
\Big\{\Gamma_x:\frac{\nu(x)}{\mu(x)}\geq c\wedge
\frac{\nu(x\leqss)}{\mu(x\leqss)}<c\ \forall\ s<\l(x)\Big\}\right)\\
& = & \sum\left\{\mu(x):\frac{\nu(x)}{\mu(x)}\geq c\wedge
\frac{\nu(x\leqss)}{\mu(x\leqss)}<c\ \forall\ s<\l(x)\right\}\\
& \leq &\sum\left\{\frac{\nu(x)}{c}:\frac{\nu(x)}{\mu(x)}\geq c\wedge
\frac{\nu(x\leqss)}{\mu(x\leqss)}<c\ \forall\ s<\l(x)\right\}\\
& = & \frac{1}{c}\sum\big\{\nu(x):\ldots\big\}
\ \leq \ \frac{1}{c},
\eqan
since $\nu$ is a (semi)measure and the set
$\left\{x\in\calX^*:\frac{\nu(x)}{\mu(x)}\geq c\wedge
\frac{\nu(x\leqss)}{\mu(x\leqss)}<c\ \forall\ s<\l(x)\right\}$
is prefix-free. Let
\beqn
B^\nu = \left\{x\ltinf:\exists\ t\geq
1\such_that\frac{w_\nu\nu(x\leqt)} {w_\mu\mu(x\leqt)}\geq
1\right\} = A_{(w_\mu/w_\nu)}^\nu,
\eeqn
then $\mu(B^\nu)\leq\frac{w_\nu}{w_\mu}$ holds. We arrange the
(semi)measures $\nu\in\calC$ in an order $\nu_1,\nu_2,\ldots$
such that the weights $w_{\nu_1},w_{\nu_2},\ldots$ are
descending. For each $c\geq 1$, we can now find an index $k$
and a set
\beqn
\calN_c = \{\nu_i:i\geq k\} \such_that \sum_{\nu\in\calN_c} w_\nu\leq\frac{w_\mu}{c}.
\eeqn
Defining $B_c=\bigcup_{\nu\in\calN^c}B^\nu$, we get
\beqn
\mu(B_c)\leq \sum_{\nu\in\calN^c}\frac{w_\nu}{w_\mu}\leq\frac{1}{c}.
\eeqn
For all $x\ltinf\notin B^\nu$, $\nu$ can never be the
maximizing element. Therefore, for all $x\ltinf\notin B_c$,
there are only finitely many $\nu\notin\calN_c$ having the
chance of becoming the maximizing element at any time. By
assumption, the maximizing element chosen from the finite set
$\calC\setminus\calN_c$ stabilizes a.s. Thus, we conclude
almost sure stabilization on the sequences in
$\calX^\infty\setminus B_c$. Since this holds for all $B_c$
and $\mu(\calX^\infty\setminus B_c)\to 1$ as $c\to\infty$, the
maximizing element stabilizes with $\mu$-probability one.
\end{Proof}

For the rest of this section, we assume that the model class
$\calC$ contains only proper measures. A measure
$\mu$ is called {\em factorizable} if there are measures $\mu_i$
on $\calX$ such that
$$ \mu(x)=\prod_{i=1}^{\l(x)} \mu_i(x_i) $$
for all $x\in\calX^*$. That is, the symbols of sequences
$x\ltinf$ generated by $\mu$ are independent. A factorizable
measure $\mu=\prod\mu_i$ is called {\em uniformly stochastic},
if there is some $\delta>0$ such that at each time $i$ the
probability of all symbols $a\in\calX$ is either 0 or at least
$\delta$. That is
\beq
\label{eq:UStochastic}
\mu_i(a)>0 \Rightarrow \mu_i(a)\geq \delta \for_all a\in\calX\und i\geq 1.
\eeq
In particular, all deterministic measures and all i.i.d.\
distributions are uniformly stochastic. Another simple example
of a uniformly stochastic measure is a probability
distribution which generates alternately random bits by fair
coin flips and the digits of the binary representation of
$\pi=3.1415\ldots$

\begin{Lemma} \label{Lemma:MDLstabilizesFinite}
Let $\mu$, $\nu$, and $\tilde\nu$ be factorizable and
uniformly stochastic measures, where $\mu$ is the true
distribution.
\\
$(i)$ The maximizing element chosen from $\mu$ and $\nu$
stabilizes almost surely.
\\
$(ii)$ If $\mu$ is not eventually always preferred over $\nu$
or $\tilde\nu$ \textup{(}in which case we the maximizing
element stabilizes a.s. by $(i)$\textup{)}, then the
maximizing element chosen from $\nu$ and $\tilde\nu$
stabilizes almost surely.
\end{Lemma}

\begin{Proof} We will show only $(ii)$, as
the proof of $(i)$ is similar but simpler. So we assume that
both $\nu$ and $\tilde\nu$ remain competitive in the process
of choosing the maximizing element, and show that then
maximizing element chosen from $\nu$ and $\tilde\nu$
stabilizes almost surely.

Let
$\nu=\prod_i\nu_i$, $\tilde\nu=\prod_i\tilde\nu_i$, and
$X_i=\frac{\tilde\nu_i(x_i)}{\nu_i(x_i)}$. The $X_i$ are
independent random variables depending on the event $x\ltinf$.
Moreover, both fractions $\frac{\nu(x\leqt)}{\mu(x\leqt)}$ and
$\frac{\tilde\nu(x\leqt)}{\mu(x\leqt)}$ are martingales
(with respect to $\mu$) and thus converge almost surely for
$t\to\infty$. We are interested only in the events in
\beqn
A_\nu=\left\{x\ltinf\in\calX^\infty:\tfrac{\nu(x\leqt)}{\mu(x\leqt)}{\rm\
converges\ to\ a\ value\ }>0\right\},
\eeqn
since otherwise $\nu$ eventually is no longer competitive. So
we assume that $\mu(A_\nu)>0$, which implies $\mu(A_\nu)=1$ by
the Kolmogorov zero-one-law (see e.g.\ \cite{Chow:88}).
Similarly, $\mu(A_{\tilde\nu})=1$ for the analogously defined
set $A_{\tilde\nu}$. That is,
\beqn
\prod_{i=1}^t X_i=\frac{\tilde\nu(x\leqt)}{\nu(x\leqt)}
=\tfrac{\tilde\nu(x\leqt)}{\mu(x\leqt)}\Big/
\tfrac{\nu(x\leqt)}{\mu(x\leqt)}
\eeqn
converges to a value $>0$ almost surely, and in particular
$0<X_i<\infty$ a.s.

Now we will use the {\it concentration function} of a real
valued random variable
$U$,
\beq \label{Eq:ConcentrationFunction}
Q(U,\eta)=\sup_{u\in\RRR}\mu(u\leq U\leq u+\eta),\ \eta\geq 0.
\eeq
This quantity was introduced by L\`{e}vy, see e.g.\
\cite{Petrov:95}. The concentration function is non-decreasing
in $\eta$. Moreover, when two independent random variables $U$
and $V$ are added, we have \cite[Lemma 1.11]{Petrov:95}
\beq \label{Eq:ConcentrationLemma}
Q(U+V,\eta)\leq\min\big\{Q(U,\eta),Q(V,\eta)\big\}\ \forall\
\eta\geq 0.
\eeq
We first assume that the following set is unbounded:
\bqa \label{Eq:ConcentrationConditionSet}
B=\left\{\sum_{i=1}^n\Big(1-Q(X_i,
\eta)\Big):n\in\NNN,\eta>0\right\}\subset\RRR^+,&& \mbox{that
is}\\
\label{Eq:ConcentrationCondition}
\sup(B)=+\infty,&&
\eqa
We show that then $\frac{\tilde\nu(x\leqt)}{\nu(x\leqt)}$
(which converges a.s.) is not concentrated in the limit. That
is, it converges to some given $c>0$, in particular to
$c=\frac{w_\nu}{w_{\tilde\nu}}$, with $\mu$-probability zero.
This shows that almost surely it does not oscillate around
$\frac{w_\nu}{w_{\tilde\nu}}$.

Define independent random variables $Y_i=\ln(X_i)$. Let
$S_n:=\sum_1^n Y_i$ and denote its almost everywhere existing
limit by $S=\sum_1^\infty Y_i$. The assertion is verified
under condition (\ref{Eq:ConcentrationCondition}), if we can
show that the distribution of $S$ is not concentrated to any
point since then also $\prod_1^\infty X_i=\exp(S)$ is not
concentrated to any point. In terms of the concentration
function defined in (\ref{Eq:ConcentrationFunction}), this
reads $Q(S,0)=0$. According to
(\ref{Eq:ConcentrationCondition}), for each $R>0$, we find
$\eta>0$ and $n\in\NNN$ such that
$\sum_{i=1}^n\big(1-Q(X_n,\eta)\big)>R$. Then, because of $X_i<\infty$
(ignoring the measure-zero set where this may fail),
\beqn
W=\max_{1\leq i\leq n} X_i=
\max\left\{\tfrac{\tilde\nu_i(x_i)}{\nu_i(x_i)}:1\leq i\leq n {\rm\ and\ }\mu(x_i)>0\right\}
\eeqn
is finite. The mapping
\beqn
\big(0,W\big]\ni w\ \mapsto\ \ln(w)\in(-\infty,\ \ln W]
\eeqn
is bijective and has derivative at least $W^{-1}$. Let
$\tilde\eta=\frac{\eta}{W}$.
Then by definition of $Y_i$, we have $Q(Y_i,\tilde\eta)\leq
Q(X_i,\eta)$ for $1\leq i\leq n$ and consequently
\beqn
\sum_{i=1}^n\Big(1-Q(Y_i,\tilde\eta)\Big)>R.
\eeqn

By the Kolmogorov-Rogozin inequality (see \cite[Theorem
2.15]{Petrov:95}), there is a constant $C$ such that
\beqn
Q(S_n,\tilde\eta)\leq
C\left(\sum_{i=1}^n\Big(1-Q(Y_i,\tilde\eta)\Big)\right)^{-\frac{1}{2}}.
\eeqn
Thus, for each $\eps>0$, we can choose $R$ sufficiently large
to guarantee $C\cdot R^{-\frac{1}{2}}<\eps$. Then
$Q(S_n,\tilde\eta)<\eps$ for $n$ and $\tilde\eta$
as before. By (\ref{Eq:ConcentrationLemma}) we conclude
\beqn
Q(S,\tilde\eta)=Q\left(S_n+\Big(\sum_{i=n+1}^\infty
Y_i\Big),\tilde\eta\right)
\leq Q(S_n,\tilde\eta)<\eps
\eeqn
and consequently $Q(S,0)=0$ since $Q$ is non-decreasing. This
proves the assertion under assumption
(\ref{Eq:ConcentrationCondition}).

Now assume that $B$ is bounded, i.e.\
(\ref{Eq:ConcentrationCondition}) does not hold. Then there is
$R>0$ such that
$\sum_{1}^n\big(1-Q(X_i,\eta)\big)\leq R$
for all $\eta>0$ and $n\in\NNN$. Since the distribution of
$X_i$ is a finite convex combination of point measures, for
each $i$ there is an
$\eta>0$ such that $Q(X_i,\eta)=Q(X_i,0)$ and thus
$\sum_{i=1}^n\big(1-Q(X_i,0)\big)\leq R$ for all $n\in\NNN$.
Therefore, also $\sum_1^\infty\big(1-Q(X_i,0)\big)\leq R$
holds. Since $\tilde\nu_i(x_i)=c_i\nu_i(x_i)$ is equivalent to
$X_i=c_i$, this implies that there are constants $c_i\geq 0$ such
that
\beq
\label{eq:sum1}
\sum_{i=1}^\infty\mu_i\big\{a:\tilde\nu_i(a)\neq c_i\nu_i(a)\big\}\leq
R.
\eeq

Next we argue that if $c_i\neq 1$ for infinitely many
$i$, then either $\nu$ or $\tilde\nu$ is eventually not competitive.
To verify this claim, let
$N_i=\big\{a:\tilde\nu_i(a)\neq c_i\nu_i(a)\big\}$ and
$M_i=\calX\setminus N_i$ and observe
that $\mu_i(N_i)<\delta$ holds for sufficiently large $i$,
since the sum (\ref{eq:sum1}) is bounded. On the other hand
$\mu$ is uniformly stochastic, so there are no events of
probability $\mu_i(a)\in(0,\delta)$, hence
$\mu_i(N_i)=0$ and $\mu_i(M_i)=1$ for
sufficiently large $i$. Now for these $i$, $c_i>1$ together
with $\nu_i(M_i)=1$ implies the contradiction
$\tilde\nu_i(M_i)=c_i>1$. So $c_i>1$ necessarily requires
$\nu_i(M_i)<1$, hence $\nu_i(M_i)\leq 1-\delta$, since $\nu$
is uniformly stochastic. If this happens infinitely often,
then $\nu$ is eventually not competitive. A symmetric argument
with $\tilde\nu$ holds for $c_i<1$.

The last paragraph shows that, if both $\nu$ and $\tilde\nu$
stay competitive, eventually $\tilde\nu_i= \nu_i$ holds a.s.
In this case, $\frac{\tilde\nu(x\leqt)}{\nu(x\leqt)}$ is
eventually constant, which completes the proof.
\end{Proof}

\begin{Cor} \label{Cor:MDLstabilizes}
Let $\calC$ be a countable class of factorizable and uniformly
stochastic measures, then the maximizing element stabilizes
almost surely.
\end{Cor}

\begin{Proof}
This follows from Theorem \ref{Theorem:MDLstabilizesF2C} and
Lemma \ref{Lemma:MDLstabilizesFinite}.
\end{Proof}

Lemma \ref{Lemma:MDLstabilizesFinite} and Corollary
\ref{Cor:MDLstabilizes} are certainly not the only or the
strongest assertions obtainable for stabilization. They rather
give a flavor how a proof can look like, even if the
distributions are not asymptotically mutually singular. On the
other hand, the given result is optimal at least in some
sense, as shown by the previous Examples
\ref{Example:NotStabilizeApprox} and
\ref{Example:NotStabilizeDependent}. In the former example,
$\mu$ is not uniformly stochastic but both $\mu$ and $\nu$ are
factorizable, while in the latter one,
$\mu$ is uniformly stochastic but $\nu$ is not factorizable.

The proof of Lemma \ref{Lemma:MDLstabilizesFinite} crucially
relies on the independence assumption, which is necessary in
order to use the Kolmogorov-Rogozin inequality. It is possible
to relax this and require independent sampling only ``every so
often". It is however not clear how to remove this condition
completely.

%%%%%%%%%%%%%%%%%%%%%%%%%%%%%%%%%%%%%%%%%%%%%%%%%%%%%%%%%%%%%%%
\section{Applications}\label{secApp}
%%%%%%%%%%%%%%%%%%%%%%%%%%%%%%%%%%%%%%%%%%%%%%%%%%%%%%%%%%%%%%%

In the following, we present some applications of the theory
developed so far. We begin by stating general loss bounds. After
that, three very general applications are discussed.

%-------------------------------
\subsection{Loss bounds}\label{subsec:Lossbounds}
%-------------------------------

So far we have only considered special loss functions, like
the square loss, the Hellinger loss, or the relative entropy.
We now show how these results, in particular the bounds for
the Hellinger loss, imply regret bounds for
\emph{arbitrary loss functions}. (As we will see, square
distance is not sufficient.) This parallels the bounds in
\cite{Hutter:02spupper,Hutter:03optisp}. The proofs are
simplified, in particular Lemma \ref{lemma:Instant2Cum}
facilitates the analysis considerably. The reader should
compare the results to the bounds for ``prediction with expert
advice", e.g.\ \cite{Cesa:97,Hutter:05expertx}.

In order to keep things simple, we restrict to binary alphabet
$\calX=\BBB$ in this section. Our results extend to general alphabet
by the techniques used in \cite{Hutter:02spupper}. Consider a
binary predictor having access to a belief probability $\ph$
depending on the current history, e.g.\
$\ph(x_t=1|x\ltt)=\frac{1}{3}$. Which actual prediction should
he output, 0 or 1? We can answer this question if we know the
\emph{loss function}, according to which losses are
assigned to the (wrong) predictions. Consider for example the
0/1 loss (also known as classification error loss), i.e.\ a
wrong prediction gives loss of 1 and a right prediction gives
no loss. Then we should predict 1 if our belief is
$\ph>\frac{1}{2}$. This may be different under other loss
functions. In general, we should predict in a \emph{Bayes
optimal} way: We should output the symbol with the least
expected loss,
\beqn
  x^\ph := \mathop{\arg\min}_{\tilde x\in\{0,1\}}
  \{(1-\ph)\ell(0,\tilde x)+\ph\ell(1,\tilde x)\},
\eeqn
where $\ell(x,\tilde x)$ is the loss incurred by prediction
$\tilde x$ if the true symbol is $x$. In the following, we
will restrict to \emph{bounded} loss functions
$\ell(x,\tilde x)\in[0,1]$. Breaking ties in the above
expression in an arbitrary deterministic way, the resulting
prediction is \emph{deterministic} for given $\ph$ and loss
function $\ell$. If $\mu$ is the true distribution as usual,
then let $l^\ph_t:=\sum_a\mu(a|x\ltt)\ell(a,x_t^\ph)$
be the $\mu$-expected loss of the
$\ph$-predictor. Then, by
\beqn
L^\ph\leqn= \Expect[l_1^\ph+...+l_n^\ph] = \sum_{t=1}^n \mu(x\ltt)l_t^\ph(x\ltt)
\eeqn
we denote the cumulative $\mu$-expected loss of the
$\ph$-predictor. With $\ph$ being the variants of the MDL predictor,
we will bound the quantity
$\Delta\leqn=L^\ph\leqn-L^\mu\leqn$, i.e.\ the cumulative
\emph{regret}, by an expression depending on
$L^\mu\leqn$ and $w_\mu^{-1}$.

We admit arbitrary non-stationary loss functions
$\ell_{x\ltt}$ which may depend on the history. Our analysis
considers the worst possible choice of loss functions and
consists of three steps. First the cumulative regret bound is
reduced to an instantaneous regret bound (Lemma
\ref{lemma:Instant2Cum}). Then the instantaneous bound is
reduced to a bound in terms of special functions of $\mu$ and
$\ph$ (Lemma \ref{lemma:Instant2Special}). Finally, the bound for
the special functions is given (Lemma \ref{lemma:Special}).

\begin{Lemma}\label{lemma:Instant2Cum}
Assume that some $\ph$-predictor satisfies the instantaneous
regret bound $\delta_t=l_t^\ph-l_t^\mu\leq
2h_t+2\sqrt{2h_t l_t^\mu}$, where
$h_t=h_t(\mu,\ph)$ is the Hellinger distance of the
instantaneous predictive probabilities (\ref{eq:Hellinger}).
Then the cumulative $\ph$-regret is bounded in the same way:
\beqn
\Delta\leqn=L^\ph\leqn-L^\mu\leqn\leq
2H\leqn(\mu,\ph)+2\sqrt{2H\leqn(\mu,\ph)L^\mu\leqn}.
\eeqn
\end{Lemma}

This and the following lemma hold with arbitrary constants,
the choices $2$ and $2\sqrt 2$ are the smallest ones for which
Lemma \ref{lemma:Special} is true. Note that if the Hellinger
distance is replaced by the relative entropy, then $2\sqrt 2$
may be replaced by $2$. Thus, normalized dynamic MDL and Bayes
mixture admit smaller bounds, compare \cite{Hutter:02spupper}.
However, this is not true for the other MDL variants, as we
have no relative entropy bound there.

\begin{Proof}
The key property is the \emph{super-additivity} of the bound.
A function $f:[0,\infty)^2\to[0,\infty)$ is said to be
super-additive if
\beqn
f(x_1+x_2,y_1+y_2)\geq f(x_1,y_1)+f(x_2,y_2).
\eeqn
The function $(H,L)\mapsto\sqrt{HL}$ satisfies this condition.
We now use an inductive argument. Assume
$\Delta^0_{2:n}\leq2H^0_{2:n}+2\sqrt{2H^0_{2:n}L^{\mu,0}_{2:n}}$, where
the summation starts at $t=2$ and the superscript $0$
indicates that the first symbol of the sequence was $0$. Let
the same hold for the first symbol $1$. Writing
$\mu_1=\mu(1|\epstr)$ and using
$\delta_1
\leq 2h_1+2\sqrt{2h_1 l^\mu_1}$, we obtain
\bqan
\lefteqn{\Delta\leqn = \delta_1+(1-\mu_1)\Delta^0_{2:n}
+\mu_1\Delta^1_{2:n}}\\
&\leq&
2\left[h_1+\sqrt{2h_1\ell_1}+
(1-\mu_1)\Big(H^0_{2:n}+\sqrt{2H^0_{2:n}L^{\mu,0}_{2:n}}\big)+
\mu_1\Big(H^1_{2:n}+\sqrt{2H^1_{2:n}L^{\mu,1}_{2:n}}\Big)\right]\\
&\leq&
2\left[H\leqn+\sqrt{2h_1\ell_1}+
\sqrt{2\Big((1-\mu_1)H^0_{2:n}+\mu_1H^1_{2:n}\Big)
\Big((1-\mu_1)L^{\mu,0}_{2:n}+\mu_1L^{\mu,1}_{2:n}\Big)}\right]\\
&\leq&2H\leqn+2\sqrt{2H\leqn L^\mu\leqn}.
\eqan
Here, the first inequality is the induction hypothesis
together with the instantaneous bound, the second bound is
Cauchy-Schwarz's inequality, and the last estimate is the
super-additivity.
\end{Proof}

\begin{Lemma}\label{lemma:Instant2Special}
Assume that some $\ph$-predictor satisfies
$\tilde\delta\leq 2h+2\sqrt{2h \tilde\ell}$ for all
$\mu,\ph\in[0,1]$, with the Hellinger distance $h=h(\mu,\ph)$ and the special
functions $\tilde\delta(\mu,\ph)$ and $\tilde\ell(\mu,\ph)$
defined in the following way, where we slightly abuse notation
and abbreviate $\mu=\mu(1|\ldots)$ and $\ph=\ph(1|\ldots)$:
\beqn
\tilde\delta=\frac{|\ph-\mu|}{\max\{\ph,1-\ph\}}\quad\und\quad
\tilde\ell=\left\{\begin{array}{l @{\quad \mbox{if}\ } l}
\mu & \mu\leq\ph\leq\frac{1}{2}, \\
\mu(1-\ph)/\ph & \mu\leq\ph \wedge \frac{1}{2}\leq\ph,\\
1-\mu & \frac{1}{2}\leq\ph\leq\mu, \\
(1-\mu)\ph/(1-\ph) & \ph\leq\mu \wedge \ph\leq\frac{1}{2}.
\end{array}\right.
\eeqn
Then for arbitrary bounded loss function
$\ell:\{0,1\}^2\to[0,1]$, we have
\beq
\label{eq:InstantBound}
\delta\leq 2h+2\sqrt{2h l^\mu}.
\eeq

\end{Lemma}

\begin{Proof}
First we show that we may assume $\ell(0,0)=\ell(1,1)=0$,
i.e.\ we do not incur loss for correct predictions. To this
end, consider the modified loss function
$\ell'(x,\tilde x)=\ell(x,\tilde x)-\ell(x,x)$ and assume w.l.o.g
$\ell'(x,\tilde x)\in[0,1]$. Then it is not hard to see that the regrets under the
original and the modified loss functions coincide, while the
expected loss of the $\mu$-predictor clearly decreases with
the modified loss function. Thus, (\ref{eq:InstantBound})
holds for $\ell$ if it holds for $\ell'$. Hence we may assume
$\ell(0,0)=\ell(1,1)=0$. For each possible outcome
$x\in\{0,1\}$, we abbreviate $\ell^x=\ell(x,1-x)$.

Now assume w.l.o.g. $\mu\leq\ph$. In order to show the
assertion, we need to consider the cases in the definition of
$\tilde\ell$ separately. We show this only for the first case,
i.e.\ $\mu\leq\ph\leq\frac{1}{2}$. Then $l^\mu=\mu\ell^1$,
$l^\ph=(1-\mu)\ell^0$. We assume that the $\mu$-predictor
outputs 0 and the $\ph$-predictor 1, otherwise they give the
same prediction and the $\ph$-predictor has no regret at all.
This condition is equivalent to $\ell^0=\ell^1\frac{u}{1-u}$
for some $u\in[\mu,\ph]$. We consider the worst case by
maximizing $l^\ph$, i.e.\ choosing $u$ as large as possible.
For this $u=\ph$, we obtain $\ell^0=\ell^1\frac{\ph}{1-\ph}$
and
\beqn
\delta=\ell^1[\tfrac{(1-\mu)\ph}{1-\ph}-\mu]=\ell^1\tilde\delta\leq
\ell^1[2h+2\sqrt{2h\tilde\ell}]\leq2h+2\sqrt{2h\ell^1\tilde\ell}\leq2h+2\sqrt{2hl^\mu},
\eeqn
showing (\ref{eq:InstantBound}) provided that
$\mu\leq\ph\leq\frac{1}{2}$. The other cases are shown
similarly.
\end{Proof}

\begin{Lemma}\label{lemma:Special}
The bound $\tilde\delta\leq 2h+2\sqrt{2h \tilde\ell}$ holds
for all $\mu,\ph\in[0,1]$, with the functions
$\tilde\delta,\tilde\ell:[0,1]^1\to[0,1]$ as defined in Lemma
\ref{lemma:Instant2Special}.
\end{Lemma}

The technical and not very interesting proof of this lemma is
omitted. The careful reader may check the assertion
numerically or graphically, as it is just the boundedness of some function on
the unit square. We remark that the bound does \emph{not} hold
if the Hellinger distance is replaced by the quadratic
distance, not even with larger constants.

\begin{Theorem}
For arbitrary non-stationary loss function which is bounded in $[0,1]$ and known to
the MDL predictors, their respective losses are bounded by
\beqn
L^{\rrho\_norm}\leqn,L^{\rrho}\leqn,
L^{\rrho^{\mathrm{static}}}\leqn,
L^{\rrho^{\mathrm{static}}\_norm}\leqn
\quad \leq \quad L^{\mu\_norm}\leqn + 2\sqrt{2cL^{\mu\_norm}\leqn\cdot w_\mu^{-1}}
+2c w_\mu^{-1},
\eeqn
where the constant $c=2, 8, 21,\mbox{ or } 32$, according to
which MDL predictor is used (compare Corollary
\ref{Cor:MuRhoAll}).
\end{Theorem}

\begin{Proof}
This follows from the above three lemmas and from
$H\leqn\leq c\!\cdot\! w_\mu^{-1}$ (Corollary~\ref{Cor:MuRhoAll}).
\end{Proof}

This shows in particular that, regardless of the loss
function, the average expected per-round regret tends to zero.
Again, the direct practical relevance of the bounds is limited
because of the potentially huge $w_\mu^{-1}$.

%-------------------------------
\subsection{Classification}
%-------------------------------

Transferring our results to pattern classification is very easy.
All we have to do is to add \emph{inputs} to our models. That is,
we consider an arbitrary input space $\calU$ and (as before) a
finite observation or output space $\calX$. A model is now a
\emph{measure}
\beqn
\nu(x|u)\in[0,1],\ x\in\calX,\ u\in\calU, \mbox{ where }
\sum_{x\in\calX} \nu(x|u)=1 \for_all u\in\calU.
\eeqn
That is, we have a distribution which is conditionalized to the
input. We restrict our discussion to measures, since there is no
motivation to consider semimeasures for classification. The
definition of a model does not include history dependence. There
is no loss of generality: We may include the history in the
arbitrary input space.

Transferring the proofs in the previous sections to the present
setup is straightforward. We therefore obtain immediately the
following corollaries.

\begin{Cor}
Let $\calC$ be a countable set of classification models containing
the true distribution $\mu$. Then for any sequence of inputs
$u\ltinf\in\calU$, we have%
\beqn
\begin{array}{l @{\quad} c @{\quad} l}
  \sum_t \Expect \sum_a \big( \mu(a|u_t)-\rrho\_norm(a|u_t,u\ltt,x\ltt) \big)^2 & \leq & 2 w_\mu^{-1}, \\[3pt]
  \sum_t \Expect \sum_a \big( \mu(a|u_t)-\rrho(a|u_t,u\ltt,x\ltt) \big)^2 & \leq & 8 w_\mu^{-1}, \\[3pt]
  \sum_t \Expect \sum_a \big( \mu(a|u_t)-\rrho^{\mathrm{static}}(a|u_t,u\ltt,x\ltt) \big)^2 & \leq & 21 w_\mu^{-1}.
\end{array}
\eeqn
(Note that although each single model formally does not depend on
the history, the MDL estimators necessarily do.)
\end{Cor}

We need not consider the normalized static variant here, since
all models are measures anyway. If there is a distribution
over $\calU$, the result therefore also holds in expectation
over the inputs. An analogue of Corollary
\ref{Cor:MDLstabilizes} is obtained as easily. If the inputs
are i.i.d., which is usually assumed for classification, then
the two conditions of factorizability and uniform
stochasticity are trivially satisfied. Therefore, the true
distribution $\mu$ is eventually discovered by MDL almost
surely. Note that in this case, the distributions are also
asymptotically mutually singular, so that the assertion also
follows from Barron's \cite{Barron:85} earlier result.

Note that again, the assumption $\mu\in\calC$ is essential. In
practical applications, if this is not clear, it may be
therefore favorable to choose a different method having
guarantees without this condition, compare
\cite{Gruenwald:04}.

%-------------------------------
\subsection{Regression}\label{subsec:Regression}
%-------------------------------

We may also apply our results in the regression setup, that is
for predicting continuous densities. Our use of the term
regression is a bit non-standard here, since it normally
refers to just estimating the mean of some prediction, where
the distribution is often assumed to be Gaussian. Again the
assumption $\mu\in\calC$ is essential, so that in practice
some other method not relying on it might be preferred.

Continuous densities cause some additional difficulties. The
observation space is now $\RRR$. This implies in particular
that, like for the loss bounds, the square distance is no
longer appropriate for our purpose\footnote{%
To see this, define a distribution
$f$ by its density
$f_n=\frac{n}{3}\chi_{[-\frac{1}{n},0]}+\frac{2n}{3}\chi_{(0,\frac{1}{n}]}$,
where $\chi$ is the characteristic function of an interval.
Let
$\tilde f(x)=f(-x)$, then the quadratic distance is $\int(f-\tilde f)^2 dx=
\frac{2n}{9}\toinfty n \infty$, whereas the relative entropy
$\int f\ln(f/\tilde f)dx=\frac{\ln 2}{3}$ is constant.
} (note that our use of the squared error is completely
different from the standard use in regression). So we will use
the Hellinger distance instead, defined similarly to
(\ref{eq:Hellinger}) by
\beq
\label{eqHellingerCont}
h(f,\tilde f) = \int \Big(
\sqrt{f(x)}-\sqrt{\tilde f(x)} \Big)^2dx
\mbox{\quad for integrable } f,\tilde f:\RRR\to[0,\infty).
\eeq
Accordingly,
$H\leqn(\mu,\ph) = \sum_t \Expect
h\big(\mu(\cdot|u_t),\ph(\cdot|u_t,u\ltt,x\ltt)\big)$ is the
cumulative Hellinger distance of two predictive distributions
$\mu$ and $\ph$. Similarly as in (\ref{eqHellingerKL}) and
(\ref{eqHellingerAbs}), the Hellinger distance is bounded by
the (continuous) relative entropy and absolute distance. This
shows in particular that the integral (\ref{eqHellingerCont})
exists.

We now consider a countable class $\calC$ of models that are
functions $\nu$ from $\calU$ to \emph{uniformly bounded
probability densities} on $\calX=\RRR$. That is, there is some
$C>0$ such that
\beq
\label{eqC}
 0\leq\nu_i(x|u)\leq C \und
\int_{-\infty}^\infty \nu_i(x|u)dx=1
\for_all i\geq 1,\ u\in\calU, \und x\in\RRR.
\eeq
for all $i\geq 1$, $u\in\calU$, and $x\in\RRR$.
The MDL estimator is then defined as the element which maximizes
the \emph{density},
$\nu^*=\arg\max_{\nu\in\calC}\{w_\nu
\nu(x\leqn|u\leqn)\}$. The uniform boundedness condition asserts that the MDL
estimator exists. It may be relaxed, provided that the MDL
estimator remains well-defined, such as for a family of Gaussian
densities which tend to the point measure.

With these definitions, the proofs of the theorems for static
and dynamic MDL can be adapted. Since the triangle inequality
holds for the Hellinger distance $\sqrt{H^2}$, we obtain the
following.

\begin{Cor}
Let $\calC$ be a countable model class according to
(\ref{eqC}), containing the true distribution $\mu$. Then for
any sequence of inputs $u\ltinf\in\calU$, we have
$H\leqn(\mu,\rrho\_norm)\leq 2 w_\mu^{-1}$,
$H\leqn(\mu,\rrho)\leq 8 w_\mu^{-1}$, and
$H\leqn(\mu,\rrho^{\mathrm{static}})\leq 21 w_\mu^{-1}$.
\end{Cor}

We may apply this for example to model classes with Gaussian
noise, concluding that the mean and the variances converge to
the true values, see \cite{Poland:05mdlreg} for an example.
It is not immediately clear how to obtain an analogue of
Corollary \ref{Cor:MDLstabilizes} for continuous densities.

%-------------------------------
\subsection{Universal Induction}\label{subsecUniversal}
%-------------------------------

Since the assertions on static and dynamic MDL have been proven
generally for \emph{semimeasures}, we may apply them to the
universal setup. Here $\calC=\calM$ is the countable set of all
lower semicomputable (= enumerable) semimeasures on $\calX^*$. So
$\calM$ contains stochastic models in general, and in particular
all models for computable deterministic sequences. There is a
one-to-one correspondence of $\calM$ to the class of all programs
on some fixed universal monotone Turing machine $U$, see e.g.
\cite{Li:97}. We will assume programs to be \emph{binary}, in
contrast to outputs, which are strings $x\in\calX^*$. This
relation defines in particular the complexities and weights of
each $\nu$ by
\beq
\label{eq:canweights}
\K w(\nu) = \mbox{length of the program for $\nu$ on $U$}, \und
w_\nu=2^{\K w(\nu)}.
\eeq
We call these weights the
\emph{canonical weights}. They satisfy
$w_\nu>0$ for all $\nu$ and $\sum_\nu w_\nu\leq 1$.

An enumerable semimeasure which dominates all other enumerable
semimeasures is called universal. The Bayes mixture
$\xi$ defined in (\ref{eq:xi}) has this property. One can show that
$\xi$ is equal within a multiplicative constant to Solomonoff's prior
\cite[Eq. (7)]{Solomonoff:64}, which is the a priori probability
that (some extension of) a string $x$ is generated provided that
the input of
$U$ consists of fair coin flips. That is
\beqn
  \xi(x)\quad\eqm\quad \M(x)\quad =\quad \sum_{\zwidths{p\ minimal:\ U(p)=x*}}
  2^{-\l(p)} \for_all x\in\calX^*.
\eeqn
Here, we use the notations
\bqan
f\leqa g:\Leftrightarrow f\leq g+O(1),
&& f\equa g:\Leftrightarrow f\leqa g\wedge g\leqa f,\\
f\leqm g:\Leftrightarrow f\leq g\cdot O(1),
&& f\eqm g:\Leftrightarrow f\leqm g\wedge g\leqm f.
\eqan

The MDL definitions in Section \ref{secMDL} directly transfer to
this setup. All bounds on the cumulative square loss (subsumed in
Corollary \ref{Cor:MuRhoAll}) therefore apply to
$\rrho=\rrho_{[\calM]}$. The necessary assumption now reads that $\mu$
must be a recursive (= computable) measure. Also, Theorem
\ref{Theorem:Solomonoff} implies Solomonoff's important universal
induction theorem.

In addition to $\calM$, we also consider the set of all recursive
measures $\tilde\calM$ together with the same canonical weights
(\ref{eq:canweights}). We define
$\tilde\xi=\xi_{[\tilde\calM]}$ and $\tilde\rrho=\rrho_{[\tilde\calM]}$.
Then $\tilde\rrho(x)\leq\tilde \xi(x)\leq\xi(x)$ and
$\rrho(x)\leq\xi(x)$ for all $x\in\calX^*$ is immediate.
It is straightforward that $\xi(x)\leqm\rrho(x)$ since
$\xi\in\calM$.
Moreover, for any string $x\in\calX^*$, define the \emph{monotone
complexity} $\Km(x)=\min\{\l(p):U(p)=x*\}$ as the length of the
shortest program such that $U$'s output starts with $x$. The
following assertion holds.

\begin{Prop}
We have $\K \tilde \rrho \geqa \Km$.
\end{Prop}

\begin{Proof}
We must show that given a string $x\in\calX^*$ and a recursive
measure $\nu$ (which in particular may be the MDL descriptor
$\nu^*(x)$) it is possible to specify a program $p$ of length at
most $\K w(\nu)+\K\nu(x)+c$ that outputs a string starting with
$x$, where constant $c$ is independent of $x$ and $\nu$.

Consider all strings $y_i\in\calX^n$ ($1\leq i\leq |\calX|^n$) of
length $n=\l(x)$ arranged in lexicographical order. Each
$y_i$ has measure $P_i=\mu(y_i)$. Let $S_i$ be the cumulated
measures: $S_0=0$ and $S_i=\sum_{k=1}^i P_k$. Let $j$ be the index
of $x$, i.e.\ $x=y_j$. Then, the interval
$[S_{j-1},S_j)\subset[0,1)$ has measure $P_j$ and therefore
contains exactly one $\lceil-\lb P_j\rceil$-bit number
$z\in[S_{j-1},S_j)$. We describe $x$ with the number $z$, this is
known as {\it arithmetic encoding} (see e.g.\ \cite{Cover:91}).
The coding is injective since $[S_{i-1},S_i)$ and $[S_{k-1},S_k)$
are disjoint for $i\neq k$.

In order to decode $z$, we may descend the $|\calX|$-ary tree of
all possible strings $y$, first considering strings of length one,
then of length two, etc. For each possible string $y$, we can
determine its binary code by approximating $\nu(x)$ sufficiently
accurately. Eventually we will find $z$, then we print the current
$y$. At this stage, $y$ might be only a prefix of $x$, since an
extension of $y$ might have a measure very close to $y$ and thus
map to the same code $z$. Therefore we continue the procedure
until all codes starting with $z$ are proper extensions of
$z$ (which may never be the case, then the algorithm
runs forever). In each step, the appropriate additional symbol is
written on the output tape. The resulting output will be $x$ or
some extension of $x$.

This algorithm can be specified in a constant $c'$ number of bits.
The description of $\nu$ needs another $\K w(\nu)$ bits. Finally,
$z$ has length $\lceil-\lb P_j\rceil\leq-\lb\nu(x)+1$. Thus, the
overall description has length $\K w(\nu)+\K\nu(x)+c$ as required.
\end{Proof}

It is also possible to prove the proposition indirectly using
\cite[Thm.4.5.4]{Li:97}. This implies that
$\Km(x)\leqa \K w(\nu)+\K\nu(x)$
for all $x\in\calX^*$ and all recursive measures
$\nu\in\tilde\calM$. Then, also
$\Km(x)\leqa \min\{\K w(\nu)+\K\nu(x)\}=\K\tilde\rrho(x)$ holds.

So together with the above observations, we have
\beq \label{Eq:ComplexityRelations}
  \Km(x)\ \leqa\ \K\tilde \rrho(x)\ \geqa \ \K\tilde \xi(x) \ \geqa\ \K \rrho(x)\ \equa\
  \K\M(x).
\eeq
On the other hand, there is a deep result in Algorithmic
information theory which states that an exact coding theorem does
\emph{not} hold on continuous sample space, $\Km(x)\stackrel{+}{\gneq}\K\M(x)$
\cite{Gacs:83}. Therefore, at least one of the above
$\geqa$ must be proper.

\begin{Problem} \label{problemGacs}
Which of the two inequalities $\K \tilde \rrho(x)\geqm\K \tilde
\xi(x)$ and $\K \tilde \xi(x)\geqm\K \rrho(x)$ is proper (or are both)?
\end{Problem}

The proof in \cite{Gacs:83} is very subtle, and the phenomenon is
still not completely understood. There is some hope that by
answering Problem \ref{problemGacs}, one arrives at a better
understanding of the continuous coding theorem and even at a
simpler proof for its failure.

%%%%%%%%%%%%%%%%%%%%%%%%%%%%%%%%%%%%%%%%%%%%%%%%%%%%%%%%%%%%%%%
\section{Discussion}\label{secDC}
%%%%%%%%%%%%%%%%%%%%%%%%%%%%%%%%%%%%%%%%%%%%%%%%%%%%%%%%%%%%%%%

In this last section, we recapitulate the main achievements of
this work and discuss their philosophical and practical
consequences. In the first place, we have shown that if
two-part MDL is used for predicting a stochastic sequence,
then the predictive probabilities converge to the true ones in
mean sum, provided that the distribution generating the
sequence is contained in the model class. The two most
important implications are almost sure convergence and loss
bounds for arbitrary loss functions.

The guaranteed convergence is slow in general: All bounds
depend linearly on $w_\mu^{-1}$, the inverse of the prior
weight of the true distribution. For large model classes, this
number must be regarded too huge to be relevant for practical
applications. Examples show that this bound is sharp. This is
in contrast to the exponentially smaller corresponding bound
for the Bayes mixture. The latter predictor however is often
computationally more expensive to approximate in practice. We
believe that this principally indicates that with MDL, some
care has to be taken when choosing the model class and the
prior. Conditions which are sufficient for fast convergence
have been given for instance in
\cite{Rissanen:96,Barron:98,Poland:04mdlspeed}. It remains a
major challenge to generalize these results in order to obtain
fast convergence under assumptions that are as weak as possible. In
particular for universal induction, this question is
interesting and possibly difficult. Even when considering only
computable Bernoulli distributions endowed with a universal
prior, fast convergence possibly holds for many environments,
but maybe not for all \cite{Poland:04mdlspeed}. We also need
to distinguish how the large error cumulates. Either the
instantaneous error remains significant for a long time, which
is critical, or the instantaneous error drops just too slowly
to be summable, e.g.\ as $O(\frac{1}{n})$, which is tolerable.
We have seen instances for both cases; compare the discussion
after Example \ref{ex:bernoulli}. In this light, the
cumulative error might not be the right quantity to assess
convergence speed.

The main results have been shown under the only assumption
that the data generating process is contained in the model
class. This condition is essential in general, as
\cite{Gruenwald:04} shows that in its absence MDL can fail
dramatically. In the universal setup, the assumption merely
requires that the data is generated in some
(probabilistically) computable way. This is a very weak
condition. Laplace, Zuse \cite{Zuse:67} and successors argue
that nature operates in a computable way, and consequently
\emph{all} thinkable data satisfies the assumption. On the
other hand, predicting with a universal model is
computationally very expensive. In particular it is provably
infeasible if the thesis of computable nature holds. Despite
these practical problems, the theory of universal prediction
is valuable since it explores the limits of computational
induction.

\textbf{Acknowledgements.} Thanks to Peter Gr\"unwald and
an anonymous reviewer for their very valuable comments and
suggestions. This work was supported by SNF grant 2100-67712.02.

%%%%%%%%%%%%%%%%%%%%%%%%%%%%%%%%%%%%%%%%%%%%%%%%%%%%%%%%%%%%%%%
%         Bibliography        %
%%%%%%%%%%%%%%%%%%%%%%%%%%%%%%%%%%%%%%%%%%%%%%%%%%%%%%%%%%%%%%%

\begin{small}
\newcommand{\etalchar}[1]{$^{#1}$}

\end{small}

\end{document}